\documentclass{article} % For LaTeX2e
\usepackage{iclr2027_conference,times}
\iclrfinalcopy

\ClearShipoutPicture

\usepackage{amsmath,amsfonts,bm}

\def\eqref#1{equation~\ref{#1}}
\def\1{\bm{1}}

\DeclareMathAlphabet{\mathsfit}{\encodingdefault}{\sfdefault}{m}{sl}
\SetMathAlphabet{\mathsfit}{bold}{\encodingdefault}{\sfdefault}{bx}{n}

\usepackage{amsmath}
\usepackage{amsthm}
\usepackage{booktabs}
\usepackage{graphicx}
\usepackage{enumitem}
\usepackage{float}
\usepackage{placeins}
\usepackage{wrapfig}
\usepackage{hyperref}
\usepackage{etoolbox}
\makeatletter
\AtBeginDocument{%
  \ifdefined\pdfrunninglinkoff
    \patchcmd{\@outputpage}
      {\setbox\@tempboxa}
      {\pdfrunninglinkoff\setbox\@tempboxa}
      {}{\PackageError{rins-layout}{Header link patch failed}{Check the output routine.}}%
    \patchcmd{\@outputpage}
      {\box\@outputbox}
      {\pdfrunninglinkon\box\@outputbox\pdfrunninglinkoff}
      {}{\PackageError{rins-layout}{Footer link patch failed}{Check the output routine.}}%
  \fi
}
\makeatother
\usepackage{url}

\usepackage{amssymb}
\usepackage{pifont}

\newcommand{\cmark}{\ding{51}}
\newcommand{\xmark}{\ding{55}}

\graphicspath{{figures/}{paper_iclr/figures/}}
\theoremstyle{plain}
\newtheorem{theorem}{Theorem}

\theoremstyle{definition}

\theoremstyle{remark}

\usepackage[framemethod=TikZ]{mdframed}
\usepackage{xcolor}

\mdfdefinestyle{theoremstyle}{
    linecolor=blue!35,
    linewidth=0.8pt,
    backgroundcolor=blue!3,
    roundcorner=2pt,
    leftmargin=0pt,
    rightmargin=0pt,
    innerleftmargin=7pt,
    innerrightmargin=7pt,
    innertopmargin=6pt,
    innerbottommargin=6pt,
    skipabove=8pt,
    skipbelow=8pt
}

\surroundwithmdframed[style=theoremstyle]{theorem}
\surroundwithmdframed[style=theoremstyle]{proposition}

\title{RINS: Residual-Image Neural Subspace \\Solvers for Large Sparse Linear Systems}

\author{
Zhongyan Ouyang$^{1,2}$ \quad
Weixin Liao$^{1,3}$ \quad
Mingquan Feng$^{3}$ \quad
Yehui Tang$^{4}$\thanks{Corresponding author: \texttt{yehuitang@ruc.edu.cn}} \quad
Junchi Yan$^{1,3}$\thanks{Corresponding author: \texttt{yanjunchi@sjtu.edu.cn}}\\
$^{1}$Shanghai Innovation Institute, Shanghai, China\\
$^{2}$Faculty of Computing, Harbin Institute of Technology, Harbin, China\\
$^{3}$School of Artificial Intelligence and School of Computer Science, \\
Shanghai Jiao Tong University, Shanghai, China\\
$^{4}$School of Smart Governance, Renmin University of China, Suzhou, China
}

\begin{document}

\maketitle

% 删除顶部 Under review 文字
\fancyhead{}
% 删除顶部横线
% \renewcommand{\headrulewidth}{0pt}

\begin{abstract}
% Accelerating solving PDE without reducing accuracy via neural network has been a significant research problem. Solving the large sparse linear systems from PDE discretizations with numerical framework maintains high accuracy but requires correction subspaces whose operator images explain the current residual for neural network to fast iterate.
Neural subspace solvers offer a promising way to accelerate the large sparse linear systems arising from PDE discretizations, but their value depends on generating effective correction directions at low computational overheads. 
We study this problem from a residual-image viewpoint: a correction subspace is useful when its image under the system operator captures the current residual.
Guided by this view, we propose Gate-RINS, which generates multiple correction directions from shared polynomial residual probes and modulates them with a lightweight residual- and coordinate-dependent pointwise gate. 
% The projected least-squares update remains unchanged, allowing the neural component to choose expansion directions 
% while the numerical closure tests them through \(\operatorname{range}(\mathbf{A}\mathbf{Q}_t)\).
% while the numerical solver determines the correction.
We also introduce a hybrid schedule that uses GRANS-style graph controllers for the first correction block and Gate-RINS thereafter, combining an expressive graph based initial step with a lower-cost corrector as the solve progress. 
Theoretical analysis establishes a unified residual-image and dynamic approximate-inverse perspective for Gate-RINS and G-RANS, showing that a single neural-generated direction can emulate the residual quality of fixed-step GMRES.
Numerical experiments on six PDE benchmarks demonstrate that Gate-RINS reaches a relative residual of $10^{-5}$ with 1.6-2.3$\times$ and 2.2-3.0$\times$ less wall-clock time than G-RANS and GMRES respectively. For Helmholtz systems with meshsize up to $10^5$, Gate-RINS also achieves the lowest residual and outperforms a GPU-accelerated parallel GMRES, while GRANS fails to reach relative residual of $10^{-4}$ in the same time budget. 
% Theoretically, we unify Gate-RINS and GRANS under a residual-image, dynamic approximate-inverse framework and show that a single neural-generated direction can emulate the residual quality of fixed-step GMRES.
\end{abstract}

\section{Introduction}\vspace{-8pt}

Solving large sparse linear systems is a fundamental computational task in PDE simulation~\citep{knoll2004jacobian}, inverse modeling~\citep{haber2014computational}, and a broad range of scientific computing~\citep{davis2006direct}. Krylov methods are widely used for these systems, with prominent examples including conjugate-gradient methods~\citep{hestenes1952methods}, Lanczos-type iterations~\citep{lanczos1950iteration}, Arnoldi/GMRES methods~\citep{arnoldi1951principle,saad1986gmres} and nonsymmetric solvers~\citep{van1992bi,freund1991qmr}. These methods construct search space through matrix-vector products and compute approximate solutions through low-dimensional projection procedures~\citep{greenbaum1997iterative,golub2013matrix,saad2003iterative,simoncini2007recent}, making them particularly attractive for large-scale sparse problems~\citep{saad2003iterative,greenbaum1997iterative,simoncini2007recent}.
% conjugate-gradient~\citep{hestenes1952methods} and Lanczos-type iterations~\citep{lanczos1950iteration} to Arnoldi/GMRES methods~\citep{arnoldi1951principle,saad1986gmres} and related nonsymmetric solvers~\citep{van1992bi,freund1991qmr}
% , from conjugate-gradient~\citep{hestenes1952methods} and Lanczos-type iterations~\citep{lanczos1950iteration} to Arnoldi/GMRES methods~\citep{arnoldi1951principle,saad1986gmres} and related nonsymmetric solvers~\citep{van1992bi,freund1991qmr}, are attractive because they use matrix-vector products and compute corrections through a projected algebraic closure~\citep{greenbaum1997iterative,golub2013matrix,saad2003iterative,simoncini2007recent}. 
%depends on whether the expanded search space contains useful correction directions at an affordable dimension. 
Their efficiency depends on both how rapidly the search space captures effective correction directions and the computational cost of constructing and maintaining that space. Classical techniques such as restarting and deflation~\citep{morgan2002gmres}, recycling~\citep{parks2006recycling}, multigrid~\citep{hackbusch2013multi,xu1992iterative} and preconditioning~\citep{benzi2002preconditioning,saad1994ilut,elman2014finite} seek to improve efficiency by controlling search-space growth or exploiting problem structure to accelerate convergence. However, their effectiveness often relies on problem-specific design and parameter choices, motivating data-driven approaches that learn to adapt solver components to the problems being solved.

% The central object is therefore the generated correction subspace. Given a residual \(\mathbf{r}_t=\mathbf{b}-\mathbf{A}\mathbf{x}_t\) and basis \(\mathbf{Q}_t\), the projected update solves $\mathbf{y}_t=\arg\min_{\mathbf{y}}\|\mathbf{A}\mathbf{Q}_t\mathbf{y}-\mathbf{r}_t\|_2$. Consequently, residual reduction is governed by the distance from \(\mathbf{r}_t\) to \(\operatorname{range}(\mathbf{A}\mathbf{Q}_t)\), not by \(\mathbf{Q}_t\) alone. Figure~\ref{fig:intro_subspace} illustrates the opportunity: enlarging the admissible family of expansion directions can yield an operator-image subspace with a smaller projected residual in the same number of iterations.

Recent advances in artificial intelligence for science (AI4Sci)~\citep{wang2023scientific,karniadakis2021physics} offer opportunities to exploit structure in PDEs through learning. Physics-informed~\citep{raissi2019physics,karniadakis2021physics} and variational neural methods~\citep{yu2018deep,kharazmi2021hp} approximate PDE solutions by optimizing differential-equation residuals or variational objectives, while neural operators~\citep{li2020fourier,lu2019deeponet,kovachki2023neural} learn solution maps across families of PDEs. Despite their promise, directly optimizing neural solution representations can be challenging for stiff or multiscale problems~\citep{wang2021understanding,krishnapriyan2021characterizing,grossmann2024can}. Moreover, a learned solution prediction alone does not ensure that a discretized system satisfies a prescribed residual tolerance~\citep{grossmann2024can,zhang2024blending}. These considerations motivate integrating learning into numerical iterations~\citep{greenfeld2019learning,zhang2024blending,liao2026grans}, where learned components can exploit problem structure while explicit residual evaluation guides solution refinement.

% Recent scientific machine learning offers another way to adapt computation to data. Physics-informed neural networks optimize differential-equation residuals directly~\citep{raissi2019physics,karniadakis2021physics}, while variational and weak-form neural PDE solvers optimize energy, Galerkin, or adversarial weak formulations~\citep{yu2018deep,sirignano2018dgm,kharazmi2021hp,jagtap2020extended,zang2020weak}. These methods can be hard to optimize on stiff or multiscale problems~\citep{wang2021understanding,krishnapriyan2021characterizing}. Operator-learning methods learn solution maps over PDE families: DeepONet learns nonlinear operators from branch--trunk representations~\citep{lu2019deeponet,lu2022comprehensive}, Fourier and multiwavelet neural operators parameterize global spectral transforms~\citep{li2020fourier,gupta2021multiwavelet,kovachki2023neural}, graph and multipole neural operators target irregular or nonlocal interactions~\citep{li2020neural,li2020multipole}, and mesh-based simulators learn dynamics on discretized physical domains~\citep{sanchez2020learning,pfaff2020learning,brandstetter2022message,takamoto2022pdebench}. These models are powerful surrogates, but their accuracy guarantees are weaker than those of conventional numerical solvers.

Within this setting, neural subspace methods learn to generate search directions while retaining a numerical projection to determine the solution update. This division assigns complementary roles to learning and numerical computation: the neural model proposes a correction space, and the projected solve selects a correction within that space. G-RANS~\citep{liao2026grans} instantiates this approach using graph attention~\cite{velivckovic2018graph} to generate residual-dependent basis vectors from matrix, coordinate, and residual information. It thereby makes subspace construction adaptive to the current solver state. However, the resulting acceleration depends on both the quality of the generated directions and their computational cost. Graph-based basis generation introduces message-passing overhead that must be offset by improved convergence. In addition, a controller trained predominantly on early residual states may become less effective as the iteration progresses and the residual is increasingly dominated by components that are difficult to eliminate.

\begin{wrapfigure}{r}{0.48\textwidth}
    \centering
    \vspace{-0.5em}
    \includegraphics[
        width=\linewidth,
        trim=125 25 75 15,
        clip
    ]{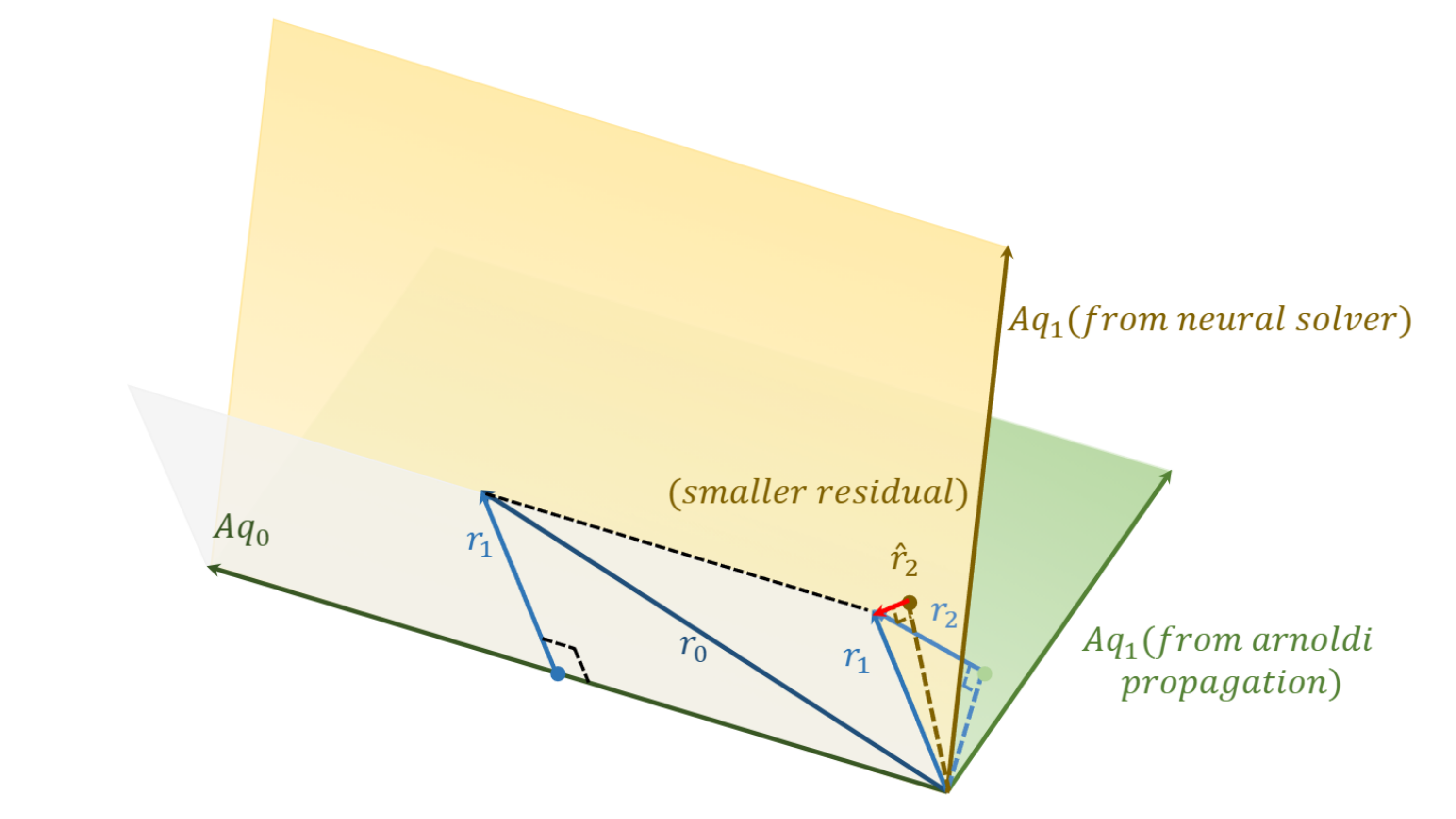}
    \caption{
    A neural solver can generate a correction space whose operator image
    leads to a smaller projected residual than an Arnoldi-generated image space.
    }
    \label{fig:intro_subspace}
    \vspace{-0.8em}
\end{wrapfigure}

To understand what makes a learned correction space effective, consider a minimum-residual update with current residual \(\mathbf{r}_t=\mathbf{b}-\mathbf{A}\mathbf{x}_t\) and correction basis \(\mathbf{Q}_t\). The update takes the form \(\mathbf{x}_{t+1}=\mathbf{x}_t+\mathbf{Q}_t\mathbf{y}_t\), where $\mathbf{y}_t\in\arg\min_{\mathbf{y}}\|\mathbf{r}_t-\mathbf{A}\mathbf{Q}_t\mathbf{y}\|_2.$ The attainable residual reduction is therefore determined by the distance from \(\mathbf{r}_t\) to the operator-image subspace \(\operatorname{range}(\mathbf{A}\mathbf{Q}_t)\). This observation provides a criterion for evaluating learned directions: their value lies in how effectively their images under \(\mathbf{A}\) represent the current residual. Fig.~\ref{fig:intro_subspace} illustrates how different correction spaces can lead to different projected residuals through their operator images. The resulting design challenge is to construct expressive, residual-dependent correction spaces at a sufficiently low cost to improve overall solution time.

Guided by this residual-image perspective, we propose Gate-RINS, a neural subspace solver that combines polynomial residual probes with lightweight pointwise modulation. The solver constructs a shared set of probes using sparse matrix-vector products and generates multiple correction directions through learnable combinations and a residual- and coordinate-dependent gate. The polynomial probes provide operator-dependent information, while the gate introduces nonlinear adaptation to the current residual. Sharing the probes across directions keeps the number of sparse operator applications tied to the probe order, rather than requiring a separate sequence of operator applications for each direction. The resulting basis is evaluated through the same projected minimum-residual update, allowing the learned component to focus on direction generation while the numerical solve determines the correction coefficients.

The changing residual structure over an iteration also motivates combining complementary basis generators. A graph-based controller can exploit matrix connectivity, coordinates, and residual information to produce effective initial corrections, whereas Gate-RINS provides a less expensive mechanism for subsequent basis generation. We therefore introduce a hybrid inference schedule that uses G-RANS for the first neural subspace block and Gate-RINS thereafter. Both controllers operate within the same projected solver, so the switch changes the basis-generation mechanism while preserving the update rule. Across six PDE-derived benchmark tasks, Gate-RINS reaches prescribed relative-residual thresholds faster than GMRES and the graph-based neural baseline in most evaluated settings, as measured by synchronized wall-clock time. Hybrid schedules provide further improvements in residual trajectories, while difficult-mode diagnostics and subspace visualizations support the residual-image interpretation of the observed gains.

% Our contributions are summarized as follows:
% \begin{itemize}[leftmargin=*, nosep]
%     \item We introduce Gate-RINS, a residual-image neural subspace solver that uses residual- and coordinate-dependent pointwise modulation to generate polynomial correction bases for sparse linear systems.

%     \item We formulate neural subspace iteration as controlled expansion and design a hybrid strategy that composes graph-based and polynomial residual-image controllers within the same projected solver.

%     \item We justify neural solver as a dynamic nonlinear approximate-inverse preconditioning and unify RINS and GRANS through this viewpoint, distinguishing diagonal-gated polynomial actions from learned sparse graph propagation.
    
%     \item Experiments show that RINS reaches fixed residual thresholds faster in synchronized wall-clock time than GMRES and a recent graph-only neural baseline~\citep{liao2026grans} in most settings, with further gains from hybrid controller schedules.
% \end{itemize}

\textbf{Our contributions are summarized as follows:}

\textbf{1) Gate-RINS.} We propose a residual-image neural subspace solver that combines shared polynomial residual probes with residual- and coordinate-dependent pointwise gating. It generates multiple correction directions while keeping polynomial operator-probe cost.

\textbf{2) Hybrid controller.} We develop a schedule that uses G-RANS for the initial correction block and Gate-RINS thereafter under the same projected minimum-residual update. This combines an expressive graph-based initial correction with lower-cost polynomial basis generation.

\textbf{3) Theoretical analysis.} We interpret the directions generated by Gate-RINS and G-RANS as residual-adaptive inverse-like actions. Separately, we show that, under the stated assumptions, a sufficiently expressive neural generator can approximate the residual quality of fixed-step GMRES with a single correction direction.

\textbf{4) Empirical results.} Across six PDE-derived tasks, Gate-RINS reaches a relative residual of \(10^{-5}\) with 1.6–2.3× and 2.2–3.0× speedups over G-RANS and GMRES, respectively, in mean wall-clock time. It also attains the lowest residual within the evaluated time budget on a separate larger Helmholtz system, while hybrid inference improves the residual trajectory on Poisson systems.

\begin{table*}[t]
\centering
\caption{
Comparison of learning-based sparse linear system solvers. "-" denotes that the neural model is not repeatedly invoked during iteration.
}
\label{tab:related_work}

\scriptsize
\setlength{\tabcolsep}{2.8pt}
\renewcommand{\arraystretch}{1.12}

\begin{tabular}{lllllll}
\toprule

Method
& SPD
& Learned Object
& Model
& Adaptive
& Extra targets
& Inference cost
\\

\midrule

\textbf{RINS (ours)}
& \textbf{\xmark}
& \textbf{Search-space basis}
& \textbf{Gated polynomial}
& \textbf{\cmark}
& \textbf{None}
& \textbf{Light-recurrent}
\\

G-RANS
& \xmark
& Search-space basis
& Graph attention
& \cmark
& None
& High-recurrent
\\

NeurKItt \citep{luo2024neurkitt}
& \xmark
& Inv. subspace
& FNO
& \xmark
& Eigenvectors
& -
\\

DCDM \citep{kaneda2023dcdm}
& \cmark
& CG direction
& CNN
& \cmark
& Rayleigh–Ritz vectors
& Medium-recurrent
\\

GNP \citep{chen2025gnp}
& \xmark
& Preconditioner
& Graph convolution
& \xmark
& Solution pairs $(\mathbf{b},\mathbf{x})$
& High-recurrent
\\

NPO \citep{li2025npo}
& \cmark
& Preconditioner
& Self-attention
& \xmark
& Solution pairs $(\mathbf{r},\mathbf{x})$
& -
\\

NeuralIF \citep{hausner2023neuralif}
& \cmark
& Preconditioner
& Coates GNN
& \xmark
& None
& -
\\

\bottomrule
\end{tabular}\vspace{-12pt}
\end{table*}

\vspace{-8pt}
\section{Relate Work}
\vspace{-8pt}
\subsection{Scientific Machine Learning for PDEs}\vspace{-8pt}

Scientific machine learning learns PDE solution fields or solution operators~\citep{karniadakis2021physics}. Physics-informed neural networks penalize differential-equation and boundary residuals~\citep{raissi2019physics}, while variational methods minimize energy functionals~\citep{yu2018deep} and deep Galerkin methods use meshfree residual sampling~\citep{sirignano2018dgm}. Training difficulties include gradient imbalance~\citep{wang2021understanding} and failures on multiscale problems~\citep{krishnapriyan2021characterizing}, motivating pseudo-time stepping in TSONN~\citep{cao2023tsonn} and sequential correction in Scale-PINN~\citep{chiu2026scale}. Operator learning instead maps PDE inputs to solution fields through DeepONet~\citep{lu2019deeponet}, Fourier neural operators~\citep{li2020fourier}, or reduced-basis-inspired representations~\citep{bhattacharya2021model}; their approximation properties are studied within a function-space framework~\citep{kovachki2023neural}. Geometry-aware extensions include graph neural operators~\citep{li2020neural}, multipole architectures~\citep{li2020multipole}, mesh-based simulators~\citep{pfaff2020learning}, and transformer-based models such as HAMLET~\citep{bryutkin2024hamlet}. Benchmarks such as PDEBench support evaluation across multiple PDE tasks~\citep{takamoto2022pdebench}, while comparisons with classical discretizations highlight remaining accuracy--cost trade-offs~\citep{grossmann2024can}. These approaches complement RINS, which learns residual-conditioned correction spaces for an assembled linear system rather than directly representing its solution field or solution operator.
\vspace{-8pt}
\subsection{Neural-Iterative Solvers}\vspace{-8pt}

Hybrid neural-numerical methods retain numerical structure while learning components such as discretizations~\citep{bar2019learning} or multigrid corrections~\citep{greenfeld2019learning}. As summarized in Table~\ref{tab:related_work}, learned linear solvers differ in what they predict and how often neural inference is used. Some learn approximate inverse operators or factorizations: NeuralIF predicts an incomplete factorization reused throughout preconditioned conjugate gradients~\citep{hausner2023neuralif}; NPO learns a matrix-dependent preconditioner incorporating multigrid structure~\citep{li2025npo}; and GNP applies a nonlinear graph-neural preconditioner recurrently within FGMRES~\citep{chen2025gnp}. Other methods modify the search space itself. DCDM predicts residual-dependent conjugate directions at each iteration~\citep{kaneda2023dcdm}, whereas NeurKItt predicts an invariant subspace once to augment subsequent Krylov iterations~\citep{luo2024neurkitt}. G-RANS generates residual-adaptive blocks of search directions~\citep{liao2026grans} using graph attention~\citep{velivckovic2018graph}, requiring recurrent graph inference followed by orthogonalization and projection. Polynomial approximation~\citep{trefethen2019approximation} and localized spectral filters~\citep{defferrard2016convolutional} provide a less expensive operator-aware alternative for basis generation. RINS retains residual-adaptive block generation and the projected minimum-residual closure, but replaces graph propagation with cached polynomial probes and lightweight pointwise modulation, amortizing a small number of SpMVs over multiple generated directions.
%%%%%%%%%%%%%%%%%%%%%%%%%%%%%%%%%%%%%%%%%%%%%%%%%%%%%%%%%%%%%%%%%%%%%%%%%%%%%%%%%%%%%%%%%%%%%
%%%%%%%%%%%%%%%%%%%%%%%%%%%%%%%%%%%%%%%%%%%%%%%%%%%%%%%%%%%%%%%%%%%%%%%%%%%%%%%%%%%%%%%%%%%%%
%%%%%%%%%%%%%%%%%%%%%%%%%%%%%%%%%%%%%%%%%%%%%%%%%%%%%%%%%%%%%%%%%%%%%%%%%%%%%%%%%%%%%%%%%%%%%
%%%%%%%%%%%%%%%%%%%%%%%%%%%%%%%%%%%%%%%%%%%%%%%%%%%%%%%%%%%%%%%%%%%%%%%%%%%%%%%%%%%%%%%%%%%%%
%%%%%%%%%%%%%%%%%%%%%%%%%%%%%%%%%%%%%%%%%%%%%%%%%%%%%%%%%%%%%%%%%%%%%%%%%%%%%%%%%%%%%%%%%%%%%
\vspace{-8pt}
\section{Methodology}\vspace{-8pt}
\label{sec:methodology}

% We use lowercase symbols for scalars, bold lowercase symbols for vectors, bold uppercase symbols for matrices, and blackboard-bold symbols for spaces and domains. The scalar \(l\) denotes a polynomial order or graph depth, \(q\) is the number of directions per graph layer, and \(m\) is the emitted block size. The transpose of each coordinate vector \(\mathbf{x}_i\) is a row of \(\mathbf{X}_{\mathrm{coord}}\), whereas \(\mathbf{x}_t\) denotes a solver iterate when indexed by iteration.

In this section, we first describe the GMRES iteration and its residual behavior, then extend fixed Krylov expansion to neural-controlled expansion of the search space and residual-image space. We use \(\mathbf{Q}_t\) for an orthonormal basis matrix, \(\mathbb{Q}_t=\operatorname{range}(\mathbf{Q}_t)\) for the corresponding search subspace, \(\mathbf{P}_t=\mathbf{Q}_t\mathbf{Q}_t^\top\) for the orthogonal projector onto \(\mathbb{Q}_t\), and \(\mathbf{P}_{\mathbf{A}\mathbf{Q}_t}\) for the orthogonal projector onto \(\operatorname{range}(\mathbf{A}\mathbf{Q}_t)\). RINS instantiates this view with a practical trade-off between SpMVs and approximation capacity: it builds polynomial probes from the current residual \(\mathbf{r}_t\), forms candidate correction directions, and, in Gate-RINS, modulates them by a pointwise MLP using local probe values and coordinates. Finally, we describe a hybrid inference framework that composes a GRANS prefix with a shared RINS-family controller for later steps under the same projected numerical closure.
\vspace{-8pt}
\subsection{Extending The Subspace Expanding Dynamics}\vspace{-8pt}
\begingroup
\setlength{\abovedisplayskip}{3pt plus 1pt minus 1pt}
\setlength{\belowdisplayskip}{3pt plus 1pt minus 1pt}
\setlength{\abovedisplayshortskip}{2pt plus 1pt minus 1pt}
\setlength{\belowdisplayshortskip}{2pt plus 1pt minus 1pt}

\textbf{Discrete Optimal Control perspective of krylov iteration.} Let \(\mathbf{A}\mathbf{x}=\mathbf{b}\), \(\mathbf{A}\in\mathbb{R}^{n\times n}\). At iteration \(t\), the residual is \(\mathbf{r}_t=\mathbf{b}-\mathbf{A}\mathbf{x}_t\). GMRES searches the affine Krylov space \(\mathbf{x}_0+\mathbb{K}_t(\mathbf{A},\mathbf{r}_0)\), where \(\mathbb{K}_t(\mathbf{A},\mathbf{r}_0)=\mathrm{span}\{\mathbf{r}_0,\mathbf{A}\mathbf{r}_0,\dots,\mathbf{A}^{t-1}\mathbf{r}_0\}\). The residual bound of GMRES at iteration step $t$ is $\|\mathbf{r}_t^{\text{GMRES}}\|_2=\min_{p\in \mathbb{P}_t,p(0)=1}\|p(\mathbf{A})\mathbf{r}_0\|_2$. Geometrically equivalently, $\|\mathbf{r}_t^{\text{GMRES}}\|_2=\|(\mathbf{I}-\mathbf{P}_{\mathbf{A}\mathbb{K}_t})\mathbf{r}_0\|_2$, where $\mathbf{P}_{\mathbf{A}\mathbb{K}_t}$ is the orthogonal projector onto the residual-image space \(\mathbf{A}\mathbb{K}_t=\operatorname{range}(\mathbf{A}\mathbf{K}_t)\) for any basis matrix \(\mathbf{K}_t\) of \(\mathbb{K}_t\); explicitly, \(\mathbf{P}_{\mathbf{A}\mathbb{K}_t}=(\mathbf{A}\mathbf{K}_t)((\mathbf{A}\mathbf{K}_t)^\top \mathbf{A}\mathbf{K}_t)^+(\mathbf{A}\mathbf{K}_t)^\top\), and superscript \(+\) denotes the Moore--Penrose pseudoinverse. Neural solvers extend this characterization to learned search subspaces (from $\mathbb{K}_t$ to $\mathbb{Q}_t$): $\|\mathbf{r}_t^{\text{NS}}\|_2=\|(\mathbf{I}-\mathbf{P}_{\mathbf{A}\mathbf{Q}_t})\mathbf{r}_0\|_2$. We view the search-space projector \(\mathbf{P}_t\) as the state of the iteration and write its dynamics as $\mathbf{P}_{t+1}=\mathbf{P}_t + \mathbf{C}_t(\mathbf{C}_t^\top \mathbf{C}_t)^+\mathbf{C}_t^\top$, where \(\mathbf{C}_t=(\mathbf{I}-\mathbf{P}_t)\mathbf{B}_t\) is the component of the newly generated block \(\mathbf{B}_t\) orthogonal to the current search subspace \(\mathbb{Q}_t\). 

\textbf{Instantiate as GMRES.} \(\mathbf{B}_t\in\mathbb{R}^{n\times1}\) is fixed for GMRES before orthogonalization by applying \(\mathbf{A}\) to the latest Arnoldi basis vector, once \(\mathbf{A}\) and \(\mathbf{r}_0\) are given. For neural solvers, \(\mathbf{B}_t\) is generated from the current residual \(\mathbf{r}_t\), the matrix \(\mathbf{A}\), optional coordinates \(\mathbf{X}_{\mathrm{coord}}\), and trained parameters \(\bm{\theta}\). It can contain \(m\) columns: \(\mathbf{B}_t\in\mathbb{R}^{n\times m}\), while requiring fewer than \(m\) expensive SpMV-equivalent propagation modules for basis generation. The parameters \(\bm{\theta}\) encode problem-family information that helps the solver generate a residual-image subspace closer to the current residual. 

\textbf{Instantiate as GRANS.} Specifically, GRANS uses a learned poly-GAT generator to construct a residual-dependent block \(\mathbf{B}_t\) and orthogonalizes it with the accumulated basis to get \(\mathbf{C}_t\): $
    \mathbf{B}_t=\operatorname{GAT}_{\bm{\theta}}(\mathbf{A},\mathbf{r}_t,\mathbf{X}_{\mathrm{coord}}),
    \mathbf{Q}_t=\operatorname{orth}\bigl([\mathbf{Q}_{t-1},\mathbf{C}_t]\bigr)=[\mathbf{Q}_{t-1},\operatorname{orth}(\mathbf{C}_t)]$.
With \(l\) message-passing layers and a \(q\)-direction head at each layer, GRANS emits \(q l\) directions per neural iteration, namely $m=ql$. This amortizes graph propagation over many directions, but each attention layer still evaluates learned edge weights, feature projections, nonlinearities, and aggregation, so it is usually more expensive than $l$ times SpMV on the same sparsity graph.

% After basis generation, GRANS uses the same minimum-residual projection as a numerical closure rather than directly predicting the next solution:
% \begin{align}
%     y_t&=\arg\min_y \|AQ_t y-r_t\|_2,
%     \qquad
%     x_{t+1}=x_t+Q_t y_t,\\
%     \|r_{t+1}\|_2
%     &=
%     \mathrm{dist}(r_t,\operatorname{range}(AQ_t)).
% \end{align}
% Thus the network generates only the correction subspace; least-squares projection determines the update.
\endgroup

%%%%%%%%%%%%%%%%%%%%%%%%%%%%%%%%%%%%%%%%%%%%%%%%%%%%%%%%%%%%%%%%%%%%%%%%%%%%%%%%%%%%%%%%%%%%%
%%%%%%%%%%%%%%%%%%%%%%%%%%%%%%%%%%%%%%%%%%%%%%%%%%%%%%%%%%%%%%%%%%%%%%%%%%%%%%%%%%%%%%%%%%%%%
%%%%%%%%%%%%%%%%%%%%%%%%%%%%%%%%%%%%%%%%%%%%%%%%%%%%%%%%%%%%%%%%%%%%%%%%%%%%%%%%%%%%%%%%%%%%%
\vspace{-5pt}
\subsection{RINS Polynomial Basis Generator}\vspace{-8pt}

RINS follows the same subspace expanding dynamics above, but specifies the neural basis block \(\mathbf{B}_t\) by a polynomial residual-image generator rather than by graph message passing. At iteration \(t\), the goal is to construct a multi-column block \(\mathbf{B}_t\in\mathbb{R}^{n\times m}\) from the current residual \(\mathbf{r}_t\), so that the orthogonalized increment \(\mathbf{C}_t=(\mathbf{I}-\mathbf{P}_t)\mathbf{B}_t\) expands \(\mathbb{Q}_t\) toward a residual image space \(\operatorname{range}(\mathbf{A}\mathbf{Q}_{t+1})\) whose projector \(\mathbf{P}_{\mathbf{A}\mathbf{Q}_{t+1}}\) leaves a smaller projected residual. The key design constraint is that \(\mathbf{B}_t\) should contain several useful directions while requiring fewer than \(m\) expensive operator-probe modules for basis generation.

\textbf{Instantiate as Static-RINS.} In the default implementation, RINS uses Chebyshev residual probes. It first scales the matrix as \(\widetilde{\mathbf{A}}=\mathbf{A}/s\), where \(s\) is a cheap maximum absolute row-sum scale. Starting from \(\mathbf{r}_t\), define $\mathbf{c}_t^{(0)}=\mathbf{r}_t, \mathbf{c}_t^{(1)}=\widetilde{\mathbf{A}} \mathbf{r}_t, \mathbf{c}_t^{(\ell)}=2\widetilde{\mathbf{A}} \mathbf{c}_t^{(\ell-1)}-\mathbf{c}_t^{(\ell-2)}, \ell=2,\ldots,l$. These vectors form a cached residual-probe dictionary $\mathbb{C}_t^{(l)}(\mathbf{A},\mathbf{r}_t)=\operatorname{span}\{\mathbf{c}_t^{(0)},\mathbf{c}_t^{(1)},\ldots,\mathbf{c}_t^{(l)}\}$. It contains \(l+1\) probe vectors and requires \(l\) SpMVs. Static-RINS chooses \(\mathbf{B}_t\) from this dictionary by learnable polynomial filters. Its \(j\)-th emitted column is
$
    \mathbf{b}_{t,j}^{\mathrm{static}}
    =
    \sum_{\ell=0}^{l}
    \alpha_{j\ell}
    \mathbf{c}_t^{(\ell)},
    j=1,\dots,m
$
with learnable coefficients \(\alpha_{j\ell}\), and
$\mathbf{B}_t^{\mathrm{static}}=[\mathbf{b}_{t,1}^{\mathrm{static}},\ldots,\mathbf{b}_{t,m}^{\mathrm{static}}] \in \mathbb{R}^{n \times m}$. The resulting basis increment is \(\mathbf{C}_t^{\mathrm{static}}=(\mathbf{I}-\mathbf{P}_t)\mathbf{B}_t^{\mathrm{static}}\), followed by QR orthogonalization and the same projected least-squares closure.

\textbf{Instantiate as Gate-RINS.} Gate-RINS enlarges this admissible set by adding a residual- and coordinate-dependent diagonal modulation before the block is appended to \(\mathbb{Q}_t\) to avoid a degeneration when $m > l + 1$. For node \(i\), let $\mathbf{z}_{t,i}=[c_{t,i}^{(0)},c_{t,i}^{(1)},\ldots,c_{t,i}^{(l)},\mathbf{x}_i],i=1,\cdots,n$. A small pointwise MLP computes $g_{t,ij}=1+s_g\tanh\!\left(\operatorname{MLP}_{\bm{\vartheta}}(\mathbf{z}_{t,i})_j\right), j=1,\ldots,m$, and the gated block is
% \begin{equation}
%     b_{t,ij}^{\mathrm{gate}}
%     =
%     g_{t,ij}
%     \sum_{\ell=0}^{l}\alpha_{j\ell}c_{t,i}^{(\ell)}.
% \end{equation}
$
    b_{t,ij}^{\mathrm{gate}}
    =
    g_{t,ij}
    \sum_{\ell=0}^{l}\alpha_{j\ell}c_{t,i}^{(\ell)}.
$
Equivalently, \(\mathbf{b}_{t,j}^{\mathrm{gate}}=\mathbf{D}_{t,j}\mathbf{b}_{t,j}^{\mathrm{static}}\), where \(\mathbf{D}_{t,j}=\operatorname{diag}(g_{t,1j},\ldots,g_{t,nj})\). Thus Gate-RINS changes the admissible block \(\mathbf{B}_t\) from a fixed polynomial span to a nodewise modulated polynomial span:
$
    \mathbf{B}_t^{\mathrm{gate}}
    =[\mathbf{D}_{t,1}\mathbf{b}_{t,1}^{\mathrm{static}},\ldots,\mathbf{D}_{t,m}\mathbf{b}_{t,m}^{\mathrm{static}}].
$
This gives the controller nonlinear capacity to alter the search-space expansion path \(\mathbb{Q}_t\mapsto\mathbb{Q}_{t+1}\), while the numerical update remains governed by the residual image projector \(\mathbf{P}_{\mathbf{A}\mathbf{Q}_{t+1}}\).

Computing the order-\(l\) dictionary requires \(l\) SpMVs. Emitting \(m\) columns of \(\mathbf{B}_t\) then reuses the cached probes, so the dominant operator-probe cost of RINS basis generation is \(l\). Gate-RINS adds only the pointwise MLP and diagonal modulation; it changes the admissible neural control family without increasing the dominant SpMV count for the probe dictionary.

%%%%%%%%%%%%%%%%%%%%%%%%%%%%%%%%%%%%%%%%%%%%%%%%%%%%%%%%%%%%%%%%%%%%%%%%%%%%%%%%%%%%%%%%%%%%%
%%%%%%%%%%%%%%%%%%%%%%%%%%%%%%%%%%%%%%%%%%%%%%%%%%%%%%%%%%%%%%%%%%%%%%%%%%%%%%%%%%%%%%%%%%%%%
%%%%%%%%%%%%%%%%%%%%%%%%%%%%%%%%%%%%%%%%%%%%%%%%%%%%%%%%%%%%%%%%%%%%%%%%%%%%%%%%%%%%%%%%%%%%%
% \vspace{-8pt}
% \subsection{Dynamic Approximate-Inverse View}\vspace{-8pt}
% \label{sec:dynamic_inverse}

\textbf{Sparse approximate inverse structure.} To better demonstrate the property discrepancy and iterative structure, we analyze these models from a unified approximate-inverse interpretation before orthogonalization. With reasonable simplification of ignoring input of coordinate, Gate-RINS equivalently emits basis preconditioned by learnable diagonal matrix: \(\mathbf{b}_{t,j}=\mathbf{D}_{t,j}p_j(\widetilde{\mathbf{A}})\mathbf{r}_t\). Under the simplification that each node is represented by a scalar feature, a single GRANS propagation layer can be equivalently interpreted as a learned one-hop sparse matrix, with its entries determined by the attention mechanism, interleaved with feature transformations and nonlinearities. Multilayer propagation enables multi-hop interactions, allowing the model to go beyond the original sparsity pattern of \(\mathbf{A}\). Block-GRANS restricts source-channel mixing to blocks, reducing source-projection parameters and restricting internal transport freedom. The minimum-residual closure selects an inverse-like correction from these generated directions, rather than requiring each direction to approximate \(\mathbf{A}^{-1}\mathbf{r}_t\) individually. Appendix~\ref{app:dynamic_inverse} formalizes this dynamic nonlinear preconditioning viewpoint and distinguishes internal channel commutativity from commutativity of the full graph operator.

\vspace{-8pt}
\subsection{Hybrid Models Bootstrap Training And Inference}\vspace{-8pt}

\textbf{Towards non-parameter-sharing control.} The preceding derivation treats the new block \(\mathbf{B}_t\) as the control that moves the search-space projector \(\mathbf{P}_t=\mathbf{Q}_t\mathbf{Q}_t^\top\). A standard neural solver uses a shared controller for all steps, namely $\mathbf{B}_t=\mathcal{B}_{\bm{\theta}}(\mathbf{A},\mathbf{r}_t,\mathbf{X}_{\mathrm{coord}}), t=0,1,\ldots$ , where $\mathcal{B}_{\bm{\theta}}$ denotes the controller and the columns of \(\mathbf{B}_t\) are newly generated search directions. However, the controlled-expansion view does not require the same control law at every step. We therefore allow a step-wise controller sequence $\mathbf{B}_t=\mathcal{B}_{\bm{\theta}_t}^{(t)}(\mathbf{A},\mathbf{r}_t,\mathbf{X}_{\mathrm{coord}}), \mathcal{B}_{\bm{\theta}_t}^{(t)}\in \{\text{GRANS},\text{Gate-RINS}_{\bm{\theta}_1},\text{Gate-RINS}_{\bm{\theta}_2},\ldots\}$. The standalone setting is the special case \(\mathcal{B}_{\bm{\theta}_t}^{(t)}=\mathcal{B}_{\bm{\theta}}\) for all \(t\). A hybrid model instead assigns controllers to solver stages, so that graph-attention and polynomial residual-image inductive biases can be used within the same projected solver.

\textbf{Multi-step bootstrap training.} Training is written in terms of frozen prefixes. Suppose controllers \(0,\ldots,t-1\) have already been trained and fixed. Starting from \(\mathbf{P}_0=0\), they generate prefix blocks \(\mathbf{B}_0,\ldots,\mathbf{B}_{t-1}\), an accumulated basis matrix \(\mathbf{Q}_t\), and the search subspace \(\mathbb{Q}_t=\operatorname{range}(\mathbf{Q}_t)\). The residual exposed to the next controller is the prefix-projected residual
$
    \mathbf{r}_t
    =
    (\mathbf{I}-\mathbf{P}_{\mathbf{A}\mathbf{Q}_t})\mathbf{r}_0,
    % \qquad
    \mathbf{P}_{\mathbf{A}\mathbf{Q}_t}
    =
    \mathbf{A}\mathbf{Q}_t\bigl((\mathbf{A}\mathbf{Q}_t)^\top \mathbf{A}\mathbf{Q}_t\bigr)^+
    (\mathbf{A}\mathbf{Q}_t)^\top .
$
The new controller \(\mathcal{B}_{\bm{\theta}_t}^{(t)}\) is trained to generate a block \(\mathbf{B}_t\) whose orthogonalized increment \(\mathbf{C}_t=(\mathbf{I}-\mathbf{P}_t)\mathbf{B}_t\) improves the next residual-image projector. Its bootstrap objective is
$
    \ell_t(\bm{\theta}_t)
    =
    \min_{\mathbf{y}}
    \left\|
    \mathbf{A}\,\operatorname{orth}([\mathbf{Q}_t,\mathbf{B}_t])\mathbf{y}-\mathbf{r}_0
    \right\|_2^2,
    % \qquad
    \mathbf{B}_t=\mathcal{B}_{\bm{\theta}_t}^{(t)}(\mathbf{A},\mathbf{r}_t,\mathbf{X}_{\mathrm{coord}}).
$
For multi-step bootstrap training, this procedure is unrolled for a short horizon, but the accumulated basis \(\mathbf{Q}_s\) and residuals are detached between steps for stability. Earlier controllers remain frozen, so the \(t\)-th model learns the residual distribution induced by the previously learned search-space trajectory rather than by the initial residual alone.

\textbf{Hybrid models inference.} This notation covers the hybrid schedules used in our experiments. In the first schedule, \(\mathcal{B}_{\bm{\theta}_0}^{(0)}\) is a GRANS controller and \(\mathcal{B}_{\bm{\theta}_{\mathrm{G}}}^{(t)}\) for \(t\ge1\) is the same trained Gate-RINS controller reused at every later step; thus inference uses GRANS for the first block and shared Gate-RINS for all later blocks. 

At inference time, standalone and hybrid models run the same cumulative projection loop. At step \(t\), the active controller emits \(\mathbf{B}_t\), the increment is orthogonalized as
$
    \mathbf{C}_t=(\mathbf{I}-\mathbf{P}_t)\mathbf{B}_t,\qquad
    \mathbf{Q}_{t+1}=[\mathbf{Q}_t, \operatorname{orth}(\mathbf{C}_t)],
    % \qquad
    \mathbf{P}_{t+1}=\mathbf{Q}_{t+1}\mathbf{Q}_{t+1}^\top ,
$
optionally with truncation if a maximum basis budget is imposed. The numerical correction is then obtained from the residual-image least-squares problem
$
\mathbf{y}_t=\arg\min_{\mathbf{y}}\|\mathbf{A}\mathbf{Q}_{t+1}\mathbf{y}-\mathbf{r}_0\|_2,
    % \qquad
    \mathbf{x}_{t+1}=\mathbf{x}_0+\mathbf{Q}_{t+1}\mathbf{y}_t,
    % \qquad
    \mathbf{r}_{t+1}=(\mathbf{I}-\mathbf{P}_{\mathbf{A}\mathbf{Q}_{t+1}})\mathbf{r}_0 .
$
Thus hybridization changes the learned control sequence \(\mathbf{B}_t\), but not the projected residual-image closure that evaluates every generated search space.

\vspace{-8pt}
\subsection{Theoretical Role of Fixed-Step GMRES Emulation}\vspace{-8pt}

% Why can a neural solver use a subspace spanned by fewer basis vectors to obtain a smaller residual? We formalize this question by asking whether the residual obtained from \(j\) neural-generated basis vectors can match the residual produced by \(m\) GMRES steps (\(j<m\)). The most stringent case is \(j=1\): a single generated correction direction is compared with the correction quality of an \(m\)-step Krylov subspace. Theorem~\ref{thm:gmres_emulation} addresses this setting. Under compact, nondegenerate assumptions on a parameterized problem family, it constructs conditions under which the \(m\)-step GMRES residual is a continuous function of the family parameter \(\bm{\mu}\), such as PDE coefficients or geometry parameters. The proof then applies a universal approximation argument to show that a continuous neural generator with one emitted direction can uniformly emulate this fixed-step GMRES residual quality. Below we state the formal result. The detailed proof is provided in Appendix~\ref{app:theory}.

Why can a neural solver achieve a smaller residual using fewer basis vectors? We formalize this by asking whether \(j\) neural-generated basis vectors can match the residual of \(m\) GMRES steps with \(j<m\). The most stringent case is \(j=1\), where a single generated direction is compared with an \(m\)-step Krylov subspace. Theorem~\ref{thm:gmres_emulation} shows that, under compact and nondegenerate assumptions on a parameterized problem family, the \(m\)-step GMRES residual varies continuously with the parameter \(\bm{\mu}\), such as PDE coefficients or geometry. A universal approximation argument then implies that a continuous neural generator emitting one direction can uniformly emulate this fixed-step GMRES residual quality. The detailed proof is given in Appendix~\ref{app:theory}.

\begin{theorem}[Single-Direction Emulation of Fixed-Step GMRES]
Let \(\mathbb{D}\) be compact and let \(\mathbf{A}(\bm{\mu})\in\mathbb{R}^{n\times n}\) and \(\mathbf{r}_0(\bm{\mu})\in\mathbb{R}^n\) be continuous in \(\bm{\mu}\in\mathbb{D}\). Assume that \(\mathbf{A}(\bm{\mu})\) is uniformly nonsingular and that, for a fixed \(m\), the matrix $\mathbf{Y}_m(\bm{\mu})=\mathbf{A}(\bm{\mu})\mathbf{K}_m(\bm{\mu}),\mathbf{K}_m(\bm{\mu})=[\mathbf{r}_0(\bm{\mu}),\mathbf{A}(\bm{\mu})\mathbf{r}_0(\bm{\mu}),\ldots,\mathbf{A}(\bm{\mu})^{m-1}\mathbf{r}_0(\bm{\mu})]$ has constant rank on \(\mathbb{D}\). Let $\rho_m^{\mathrm{GMRES}}(\bm{\mu})=\min_{\mathbf{y}} \|\mathbf{r}_0(\bm{\mu})-\mathbf{Y}_m(\bm{\mu})\mathbf{y}\|_2=\|\mathbf{r}_0(\bm{\mu})-\mathbf{Y}_m(\bm{\mu})\mathbf{y}_m^*\|_2$, where $\mathbf{y}_m^*=\mathbf{Y}_m(\bm{\mu})^+\mathbf{r}_0$.
For every \(\varepsilon>0\), there exists a continuous neural generator \(\Phi_{\bm{\theta}}\) that maps the system data \((\operatorname{vec}\mathbf{A}(\bm{\mu}),\mathbf{r}_0(\bm{\mu}))\) to a correction vector \(\bm{\phi}_{\bm{\theta}}(\bm{\mu})\in\mathbb{R}^n\) such that the one-dimensional projected update over \(\operatorname{span}\{\bm{\phi}_{\bm{\theta}}(\bm{\mu})\}\) satisfies
$
    \min_{\alpha\in\mathbb{R}}
    \|\mathbf{r}_0(\bm{\mu})-\alpha \mathbf{A}(\bm{\mu})\bm{\phi}_{\bm{\theta}}(\bm{\mu})\|_2
    \le
    \rho_m^{\mathrm{GMRES}}(\bm{\mu})+\varepsilon,
    \forall \bm{\mu}\in\mathbb{D} .
$
\end{theorem}

\vspace{-8pt}
\section{Experiments}\vspace{-8pt}

We evaluate residual-reduction efficiency, and examine large-scale behavior, correction trajectories, spectral alignment and single-action approximate-inverse quality. Hybrid schedules and scale/geometry transfer tests further probe how the generator architecture affects solver behavior.

\vspace{-8pt}
\subsection{Main Benchmark}\vspace{-8pt}

The main benchmark tests whether learned residual-adaptive basis generation improves residual-reduction efficiency over GMRES under a shared evaluation protocol.

\textbf{Tasks and solvers.} We evaluate RINS, GRANS, and GMRES on six PDE-induced sparse linear-system tasks: Poisson, Helmholtz, heat, advection-diffusion, Klein--Gordon, and wave; Appendix~\ref{app:experimental_details} gives the discretizations. Unless otherwise specified, RINS denotes Gate-RINS and Static-RINS denotes the ungated polynomial-filter ablation. A separate Helmholtz solve at \(n\approx100000\) is reported in Section~\ref{sec:large_scale_comparison} and Figure~\ref{fig:experiment_summary_row}a. RINS and GRANS use the same projected minimum-residual update, with 200 neural iterations and 20 new basis vectors per iteration. The \(n\approx1000\) timing table uses CPU-optimized GMRES, which runs until the minimum neural-solver residual is reached; GPU GMRES was not faster at this scale in our measurements.

\textbf{Statistics and hardware.} Residual curves use a fixed training seed and report the median over 10 held-out samples, with shaded sample interquartile ranges (25th--75th percentiles). Timing tables pool all valid seed/sample time-to-threshold observations and report their pooled mean and standard deviation. No external preconditioner is used in this benchmark. Appendix~\ref{app:workflow} reports separate integration experiments in which both training and inference use preconditioned data, demonstrating additional acceleration from combining RINS with preconditioning. All GPU neural-solver experiments use one NVIDIA H100 GPU.

\begin{table}[!tbp]
\centering
\caption{Time in seconds to reach relative residual $10^{-5}$ at \(n\approx1000\). Solver-time entries are the pooled mean and standard deviation over all valid seed/sample observations. The last two columns report the speedup of RINS over GRANS and GMRES, computed from pooled mean times.}

\label{tab:main_benchmark}
% Match Table 1: the ICLR style sets both small and footnotesize to 9pt.
\scriptsize
\setlength{\tabcolsep}{2pt}
\begin{tabular*}{\linewidth}{@{\extracolsep{\fill}}lccccc@{}}
\toprule
PDE & RINS & GRANS & GMRES & GRANS & GMRES \\
\midrule
Poisson             & $0.105\pm0.013$ & $0.219\pm0.041$ & $0.292\pm0.035$ & 2.09x & 2.79x \\
Helmholtz           & $0.121\pm0.010$ & $0.194\pm0.036$ & $0.335\pm0.037$ & 1.61x & 2.77x \\
Heat                & $0.058\pm0.004$ & $0.136\pm0.025$ & $0.128\pm0.019$ & 2.34x & 2.20x \\
Advection-Diffusion & $0.105\pm0.007$ & $0.232\pm0.024$ & $0.320\pm0.032$ & 2.20x & 3.04x \\
Klein-Gordon        & $0.061\pm0.009$ & $0.120\pm0.014$ & $0.138\pm0.045$ & 1.97x & 2.27x \\
Wave                & $0.061\pm0.008$ & $0.130\pm0.031$ & $0.146\pm0.045$ & 2.15x & 2.41x \\
\bottomrule
\end{tabular*}
\vspace{-5pt}
\end{table}

\begin{figure}[!tb]
    \centering
    \includegraphics[width=0.92\textwidth, trim=5 5 5 5, clip]{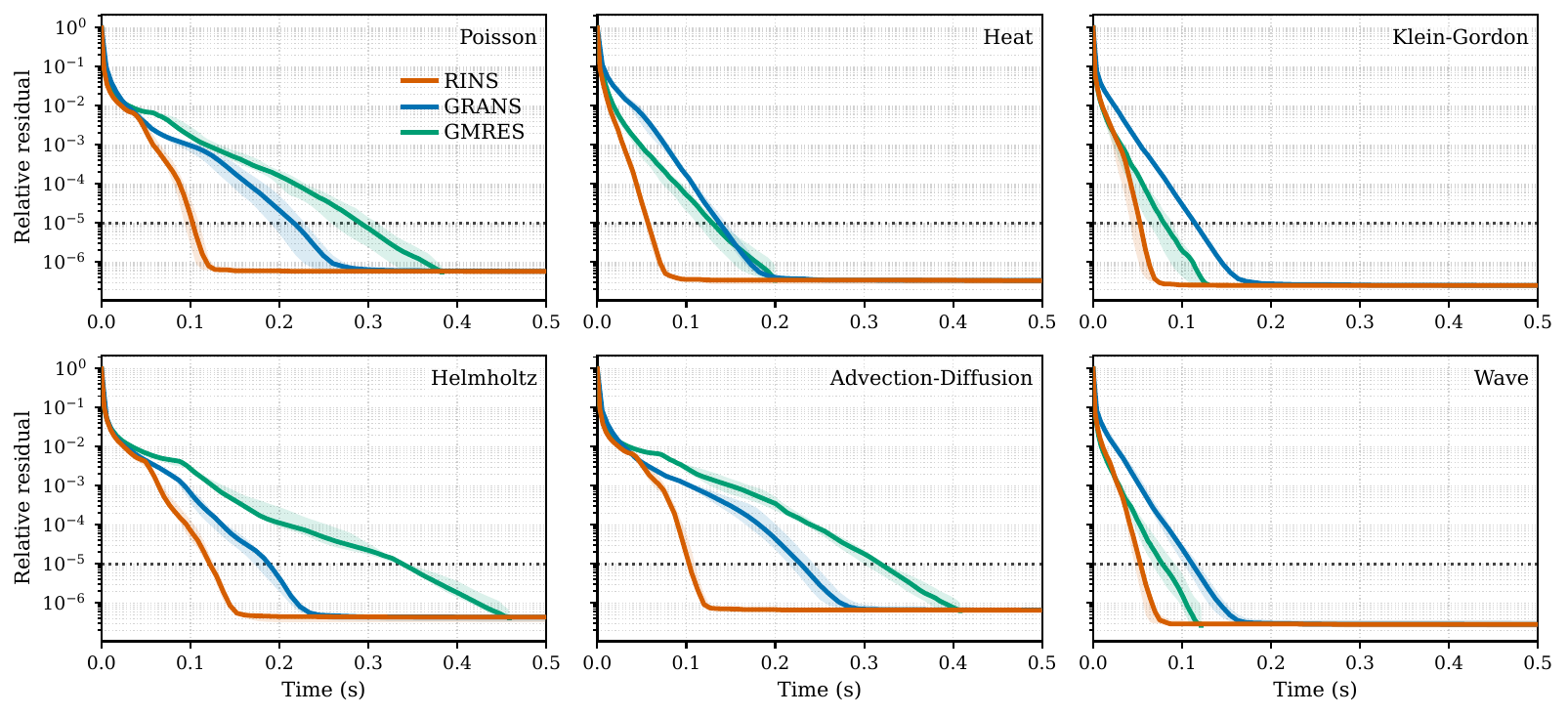}
    \vspace{-12pt}
    \caption{Main benchmark relative-residual curves at $n\approx1000$. GMRES is run until it reaches $\min(\mathrm{RelRes}_{\mathrm{RINS}},\mathrm{RelRes}_{\mathrm{GRANS}})$. Curves use one fixed training seed and report median residual over 10 held-out samples, with shaded bands showing the sample interquartile range.}
    \vspace{-12pt}
    \label{fig:main_benchmark_n1000}
\end{figure}

Figure~\ref{fig:main_benchmark_n1000} and Table~\ref{tab:main_benchmark} show the advantage of RINS over GRANS and GMRES in rapidly reducing the residual across the different linear systems at \(n\approx1000\). Because RINS and GRANS add 20 directions at each neural iteration, their subspace dimension grows faster per neural step than the GMRES Krylov dimension grows per Arnoldi step. For RINS this larger basis block does not imply a proportional increase in sparse matrix-vector products: an order-$L$ RINS call constructs the shared polynomial dictionary with $L$ SpMVs and then emits the 20 basis vectors through inexpensive linear combinations. This amortization helps explain why RINS can reach the residual threshold earlier in wall-clock time even though a single neural block iteration contains more work than one Arnoldi step. GRANS also amortizes \(L\) message-passing layers over a larger \(K\times L\) emitted basis block, but each graph-attention propagation layer is usually more expensive than one SpMV because it computes learned edge weights and feature transforms before aggregation. Under the measured synchronized wall-clock protocol, the neural solvers generally reach the target relative-residual level sooner. Appendix~\ref{app:one_basis} further shows that the polynomial residual-image generator is not solely benefiting from multi-vector batching. Restricted to generating one basis direction per neural iteration is faster as well.

\vspace{-8pt}
\subsection{Large Scale Comparison}\vspace{-8pt}
\label{sec:large_scale_comparison}

\begin{figure}[!tb]
    \centering
    \begin{minipage}[t]{0.32\textwidth}
        \centering
        \IfFileExists{figures/speed_test_single.pdf}{%
            \includegraphics[width=\linewidth]{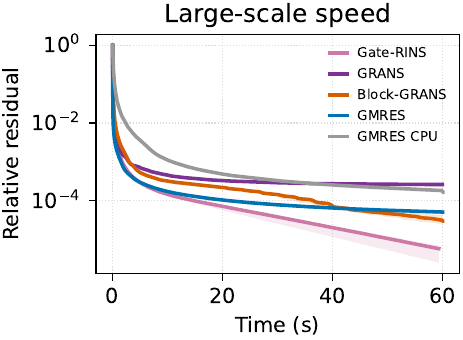}%
        }{%
            \fbox{\parbox[c][0.17\textheight][c]{0.9\linewidth}{\centering Speed-test panel generated from cached CSV rows.}}%
        }\\[-0.4em]
        {\scriptsize (a) Helmholtz speed, \(n\approx 10^5\)}
    \end{minipage}\hfill
    \begin{minipage}[t]{0.32\textwidth}
        \centering
        \IfFileExists{figures/pca_trajectory_single.pdf}{%
            \includegraphics[width=\linewidth]{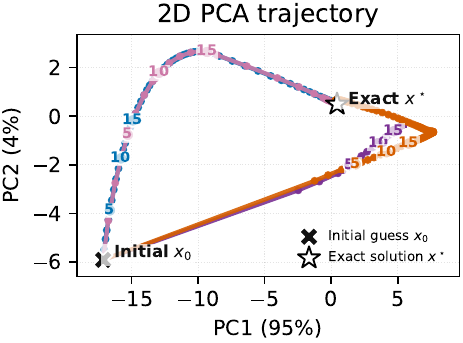}%
        }{%
            \fbox{\parbox[c][0.17\textheight][c]{0.9\linewidth}{\centering PCA trajectory placeholder.}}%
        }\\[-0.4em]
        {\scriptsize (b) PCA trajectory}
    \end{minipage}\hfill
    \begin{minipage}[t]{0.32\textwidth}
        \centering
        \IfFileExists{figures/difficult_mode_aq_overlap_with_summary.pdf}{%
            \includegraphics[width=\linewidth]{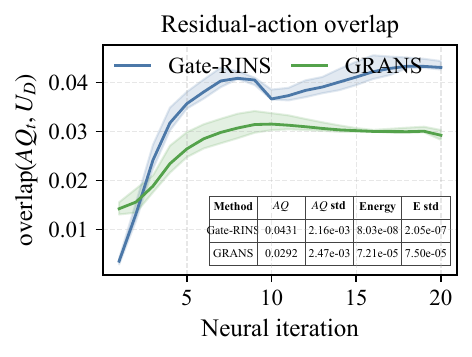}%
        }{%
            \fbox{\parbox[c][0.17\textheight][c]{0.9\linewidth}{\centering Residual-action overlap placeholder.}}%
        }\\[-0.4em]
        {\scriptsize (c) Residual-action overlap}
    \end{minipage}
    \vspace{-10pt}
    \caption{Compact experimental summary panels. Panel (a) uses GPU GMRES for the large-scale Helmholtz comparison.}
    \vspace{-12pt}
    \label{fig:experiment_summary_row}
\end{figure}

Figure~\ref{fig:experiment_summary_row}a focuses on a large-scale Helmholtz instance with \(n\approx 10^5\) unknowns. It reports relative residual versus cumulative solve time for Gate-RINS, GRANS, Block-GRANS, and GPU GMRES, using the same synchronized H100 timing convention as the main benchmark. Block-GRANS is the structured GRANS variant described in Appendix~\ref{app:block_grans_arch}. The blue GMRES curve in this large-scale panel is the GPU implementation. This panel tests whether the residual-image polynomial generator remains useful when the sparse operator and accumulated projection basis are substantially larger than in the main \(n\approx1000\) table. The resulting curve summarizes the practical large-scale trade-off: a lower residual trajectory is meaningful only when it survives the full wall-clock cost of basis generation, orthogonalization, and projected least squares. In large-scale tasks, RINS achieves the fastest residual reduction and the lowest final residual within the same computational budget, followed by Block-GRANS, demonstrating favorable scalability.

\vspace{-8pt}
\subsection{Trajectory For Different Models}\vspace{-8pt}

Figure~\ref{fig:experiment_summary_row}b visualizes solver iterates by projecting the trajectories of different methods onto a shared two-dimensional PCA plane. The panel shows how the iterates move from the initial guess toward the exact solution and reveals that different neural solvers exhibit different paths at a different speed. Gate-RINS stays close to the GMRES trajectory, consistent with its polynomial residual-probe construction, but each neural step can add a richer block of directions and therefore achieves larger residual reduction after the projected update. By contrast, GRANS and Block-GRANS follow visibly different paths, reflecting the additional freedom introduced by learned graph attention and, for Block-GRANS, its structured feature-block message projection. The colored dotted trajectories use the same method-color assignment as Figure~\ref{fig:experiment_summary_row}a, and the numeric markers indicate the projected locations of the solution vectors at the corresponding neural or Krylov iteration steps. The legend labels only the initial guess and exact solution.

\vspace{-8pt}
\subsection{Difficult-Mode Spectral Diagnostics}\vspace{-8pt}

Figure~\ref{fig:experiment_summary_row}c asks whether the residual-image space \(\mathrm{range}(\mathbf{A}\mathbf{Q}_t)\) covers spectral components that are hard for the current residual to remove. We run this diagnostic on Poisson systems, where the matrix is symmetric positive definite after discretization. For each test matrix, we form \(\mathbf{A}=\mathbf{V}\bm{\Lambda} \mathbf{V}^\top\), sort eigenvectors by increasing eigenvalue magnitude, and define \(\mathbf{U}_{\mathrm{D}}=[\mathbf{v}_{i_1},\ldots,\mathbf{v}_{i_s}]\) with \(s=8\). At diagnostic step \(t\), after the tested solver emits \(\mathbf{B}_t\) and updates \(\mathbf{Q}_t\), we measure
% $
%     o_{\mathbf{A}\mathbf{Q}}(t)=\frac{\|\widehat{\mathbf{A}\mathbf{Q}_t}^{\top}\mathbf{U}_{\mathrm{D}}\|_{\mathrm{F}}}
%     {\sqrt{\dim(\mathbf{A}\mathbf{Q}_t)\dim(\mathbf{U}_{\mathrm{D}})}},
%     % \qquad.
% $
$
    o_{\mathbf{A}\mathbf{Q}}(t)=\|\widehat{\mathbf{A}\mathbf{Q}_t}^{\top}\mathbf{U}_{\mathrm{D}}\|_{\mathrm{F}} 
    /
    {\sqrt{\dim(\mathbf{A}\mathbf{Q}_t)\dim(\mathbf{U}_{\mathrm{D}})}},
    % \qquad.
$
where $\widehat{\mathbf{A}\mathbf{Q}_t}=\operatorname{orth}(\mathbf{A}\mathbf{Q}_t)$. Larger values indicate stronger alignment between the residual space used by the projected least-squares update and the difficult eigenspace. Except for the first few iterations, Gate-RINS maintains stronger overlap with these difficult modes, suggesting that its gated polynomial probes become especially useful after the early residual components have already been removed. This alignment-based observation also motivates the later hybrid schedule: an expressive graph controller can provide the initial correction, while Gate-RINS can take over when the solve enters the difficult-mode-dominated stage. The shaded bands are the 25th--75th percentile ranges over the 10 held-out Poisson samples for one fixed training seed, and the inset table reports the final \(\mathbf{A}\mathbf{Q}\)-overlap statistics.

\vspace{-8pt}
\subsection{Hybrid Inference Acceleration}\vspace{-8pt}

\begin{figure}[!htbp]
    \centering
    \includegraphics[width=\textwidth]{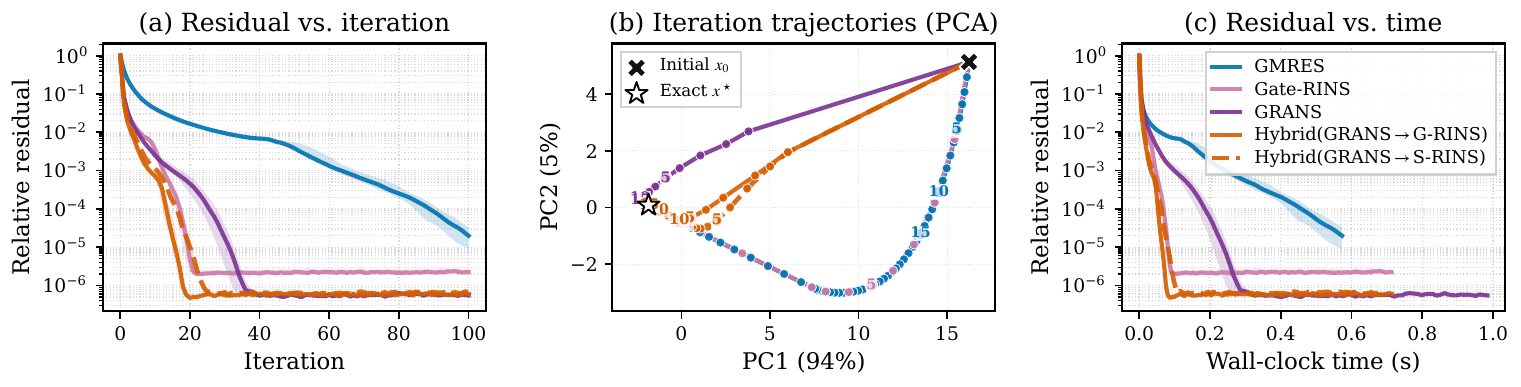}
    \vspace{-10pt}
    \caption{Hybrid inference acceleration on the Poisson problem. }
    \vspace{-12pt}
    \label{fig:hybrid_inference_poisson}
\end{figure}

From the perspective of discrete optimal control, the hybrid controller enables non-parameter-sharing, or further, non-model-sharing control at different iteration steps. Figure~\ref{fig:hybrid_inference_poisson} evaluates the hybrid controller described in Section~\ref{sec:methodology}: the first neural subspace block is generated by a GRANS prefix trained specifically for the first bootstrap step of this hybrid experiment, and all subsequent blocks are generated by a shared trained RINS-family controller under the same projected least-squares closure. This GRANS prefix is frozen after its first-step training and is not the standalone GRANS checkpoint used elsewhere. We compare two schedules with the same frozen GRANS prefix: Hybrid(GRANS$\to$G-RINS), which uses Gate-RINS after the first step, and Hybrid(GRANS$\to$S-RINS), which uses Static-RINS (no pointwise gate) with the same emitted basis count \(m\) and polynomial order \(L\). On the Poisson test systems, the mixed schedules test whether the graph controller's initial correction can be composed with either gated or ungated polynomial residual-image generators. The wall-clock panel shows that this improvement is not only an iteration-count effect: because only the first step uses graph message passing, most inference steps keep the RINS cost profile while benefiting from a better initialized subspace trajectory. 

% The architectural significance is that hybrid inference decouples the numerical closure from the learned controller family. Once each step returns a candidate block \(\mathbf{B}_t\), the solver applies the same orthogonalization and residual-image projection regardless of whether \(\mathbf{B}_t\) came from graph attention or from polynomial probes. This makes the method modular: expensive, expressive controllers can be allocated to stages where their inductive bias is most useful, while cheaper Gate-RINS controllers handle the remaining iterations. In this sense, hybridization is a controlled subspace-expansion policy that composes different learned search-space generators inside one transparent iterative solver.

% \vspace{-8pt}
\subsection{Out-of-Distribution Generalization}\vspace{-8pt}
\label{sec:ood_generalization}

\begin{figure}[!htbp]
    \centering
    \begin{minipage}[t]{0.32\textwidth}
        \centering
        \IfFileExists{figures/exp2-1_iters_vs_size.pdf}{%
            \includegraphics[width=\linewidth]{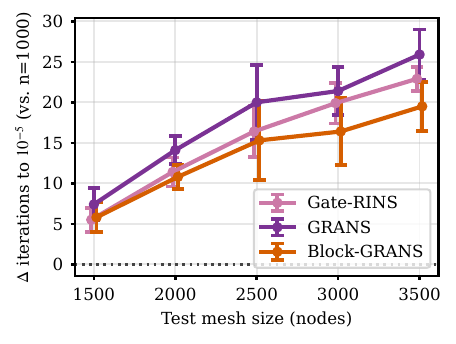}%
        }{%
            \fbox{\parbox[c][0.13\textheight][c]{0.9\linewidth}{\centering Scale-up generalization placeholder.}}%
        }\\[-0.4em]
        {\scriptsize (a) Scale-up, trained \(\approx1000\)}
    \end{minipage}\hfill
    \begin{minipage}[t]{0.32\textwidth}
        \centering
        \IfFileExists{figures/exp2-2_iters_vs_size.pdf}{%
            \includegraphics[width=\linewidth]{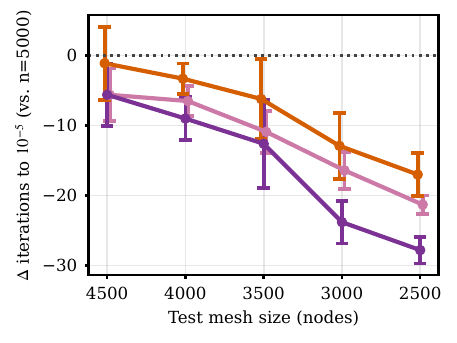}%
        }{%
            \fbox{\parbox[c][0.13\textheight][c]{0.9\linewidth}{\centering Scale-down generalization placeholder.}}%
        }\\[-0.4em]
        {\scriptsize (b) Scale-down, trained \(\approx5000\)}
    \end{minipage}\hfill
    \begin{minipage}[t]{0.32\textwidth}
        \centering
        \IfFileExists{figures/exp3_residual_iteration_ishape.pdf}{%
            \includegraphics[width=\linewidth]{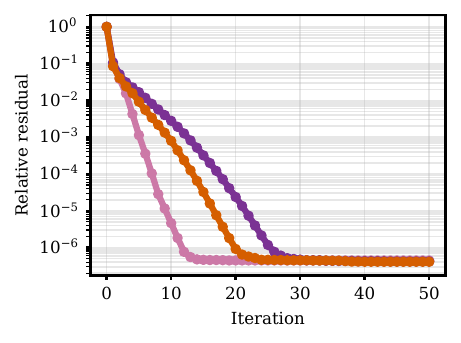}%
        }{%
            \fbox{\parbox[c][0.13\textheight][c]{0.9\linewidth}{\centering Geometry generalization placeholder.}}%
        }\\[-0.4em]
        {\scriptsize (c) Geometry, I-shaped domain}
    \end{minipage}
    \vspace{-9pt}
    \caption{Out-of-distribution generalization on Poisson systems under \(\pm30\%\) parameter variation, with a shared method-color legend across the three panels (\textcolor[HTML]{CC79A7}{Gate-RINS}, \textcolor[HTML]{7B3294}{GRANS}, \textcolor[HTML]{D55E00}{Block-GRANS}).}
    \vspace{-9pt}
    \label{fig:exp_generation}
\end{figure}

We evaluate Gate-RINS, GRANS, and Block-GRANS under scale and geometry shifts with a fixed training seed and \(\pm30\%\) parameter variation. Figure~\ref{fig:exp_generation}a tests controllers trained at approximately 1000 nodes on larger meshes (1500--3500 nodes), while panel (b) tests controllers trained at approximately 5000 nodes on smaller meshes (2500--4500 nodes). Both panels report the change in mean neural iterations to relative residual \(10^{-5}\) from the training-scale baseline, with sample-standard-deviation error bars over 10 test systems. Across the tested scale ranges, GRANS shows the largest overall iteration increase as meshes grow and the largest decrease as meshes shrink. By contrast, Block-GRANS shows the smallest overall increase and decrease, consistent with a more restricted transport family, though not establishing a causal role for commutativity (Appendix~\ref{app:dynamic_inverse}). Gate-RINS nevertheless requires fewer iterations than GRANS at every tested size (Appendix~\ref{app:grid_generalization}). Panel (c) tests transfer from rectangular Poisson domains to an I-shaped domain: all three methods converge, with Gate-RINS showing the fastest residual reduction in this test. Appendix~\ref{app:geometry_generalization} details the geometry protocol.

\vspace{-6pt}
\subsection{Single-Action Approximate-Inverse Quality}\vspace{-8pt}
\label{sec:inverse_action}

% We test the viewpoint in Appendix~\ref{app:dynamic_inverse} on 10 Poisson systems at $n\approx1000$, with $\mathbf{x}_0=0$ and $\mathbf{r}=\mathbf{b}$. For each system $s$, we compute $e_s=\|\mathbf{A}_s\mathbf{z}_s-\mathbf{b}_s\|_2/\|\mathbf{b}_s\|_2$ on the original full system, then report the arithmetic mean and sample standard deviation across systems. All three neural solvers use one-basis checkpoints: one forward call emits a single vector $\mathbf{q}$ and the correction is $\mathbf{z}=\mathbf{q}(\mathbf{A}\mathbf{q})^+\mathbf{b}$, where $+$ denotes the Moore--Penrose pseudoinverse. NeuralIF learns factors on 200 training systems' SPD interior blocks and restores the Dirichlet boundary values exactly; GNP is trained separately for each test matrix and is therefore not a held-out-matrix generalization baseline. ILU uses fixed drop-tolerance/fill-factor pairs.

\setlength{\intextsep}{2pt}
\setlength{\columnsep}{10pt}

\begin{wraptable}{r}{0.3\textwidth}
\vspace{-15pt}
\centering
\caption{\footnotesize Single-action relative residual: mean (sample standard deviation). ILU labels specify drop tolerance and fill factor.}
\label{tab:inverse_action}
\scriptsize
\setlength{\tabcolsep}{3pt}

\begin{tabular*}{\linewidth}{@{\extracolsep{\fill}}lc@{}}
\toprule
Method & Mean (Std) \\
\midrule
NeuralIF           & 0.20848 (0.02067) \\
GNP                & 0.14405 (0.04038) \\
ILU ($10^{-2},2$)  & 0.28199 (0.02387) \\
ILU ($10^{-4},10$) & 0.00183 (0.00019) \\
Gate-RINS          & 0.12817 (0.01473) \\
GRANS              & 0.10485 (0.03650) \\
Block-GRANS        & 0.08147 (0.01419) \\
\bottomrule
\end{tabular*}

% \vspace{-10pt}
\end{wraptable}

% Table~\ref{tab:inverse_action} shows that the single-direction neural corrections attain lower mean residuals than NeuralIF, GNP, and the coarser ILU, supporting residual-conditioned inverse-action approximation even in a one-dimensional correction space. The higher-fill ILU is included only as an accuracy reference, not a cost-matched competitor. Among neural solvers, Block-GRANS gives the lowest mean residual, followed by GRANS and Gate-RINS. Gate-RINS thus trades single-action quality for lightweight polynomial probing; the separate hybrid experiment in Figure~\ref{fig:hybrid_inference_poisson} exploits a graph-based initial correction followed by inexpensive Gate-RINS steps. Its distinct checkpoints and multi-direction schedule mean that this single-action test motivates the design, rather than establishing hybrid gains or a uniform approximate-inverse guarantee.

We test the approximate inverse viewpoint in Appendix~\ref{app:dynamic_inverse} on 10 Poisson systems at $n\approx1000$, with $\mathbf{x}_0=0$ and $\mathbf{r}=\mathbf{b}$. For each system, we report the relative residual $e_s=\|\mathbf{A}_s\mathbf{z}_s-\mathbf{b}_s\|_2/\|\mathbf{b}_s\|_2$ and summarize its mean and sample standard deviation. Each neural solver uses a one-basis checkpoint, producing one vector $\mathbf{q}$ with correction $\mathbf{z}=\mathbf{q}(\mathbf{A}\mathbf{q})^+\mathbf{b}$. $z=\mathbf{M}\mathbf{b}$ for precondition methods, and $\mathbf{M}$ is the approximate inverse. NeuralIF and neural solvers are trained on 200 systems, GNP separately on each test matrix, and ILU uses fixed drop-tolerance/fill-factor pairs. Table~\ref{tab:inverse_action} shows that Block-GRANS, GRANS, and Gate-RINS achieve lower mean residuals than NeuralIF, GNP, and the coarser ILU. Among neural methods, Block-GRANS performs best, followed by GRANS and Gate-RINS. Gate-RINS prioritizes lightweight polynomial probing. This also validates the hybrid experiment in Figure~\ref{fig:hybrid_inference_poisson}.

\vspace{-8pt}
\section{Conclusion and Outlook}\vspace{-8pt}

We presented RINS, a residual-image neural subspace solver that preserves projected minimum-residual updates while learning only the basis generator. Gate-RINS combines residual-image polynomial probes with a lightweight pointwise gate, improving subspace flexibility without increasing the dominant SpMV cost. Experiments show that faster convergence correlates with better alignment between $\mathrm{range}(\mathbf{A}\mathbf{Q}_t)$ and difficult residual components. Despite higher per-block cost than a GMRES Arnoldi step, RINS achieves larger residual reduction per iteration and lower end-to-end runtime on the evaluated PDE systems. Future work will explore new neural-solver variants and matrix-free deployment for large-scale scientific computing.

\clearpage
\subsection*{AI use statement}\vspace{-8pt}
In this work, we used generative AI tools for drafting assistance and editing support during manuscript preparation. We have reviewed all AI-assisted work and take responsibility for the final content of this work, including text, claims, and artifacts produced with the aid of generative AI.

\subsection*{Ethics statement}\vspace{-8pt}
This work studies numerical solvers for sparse linear systems arising from PDE discretizations. We do not use human-subject data or personally identifiable information. Potential risks are indirect and relate to downstream scientific-computing applications using the proposed solver.

\subsection*{Reproducibility statement}\vspace{-8pt}
We describe the model architecture, training schedule, benchmark construction, evaluation protocol, timing convention, and ablation settings in the main text and appendix. The experiments use fixed train/test splits and report median trajectories with uncertainty bands over held-out systems.

\bibliography{iclr2027_conference}
\bibliographystyle{iclr2027_conference}

\appendix

\clearpage
\section*{Appendix Contents}
\noindent\hyperref[app:theory]{\ref*{app:theory}\quad Uniform Emulation of Fixed-Step GMRES}\dotfill\pageref{app:theory}\par
\noindent\hyperref[app:experimental_details]{\ref*{app:experimental_details}\quad Experimental Details}\dotfill\pageref{app:experimental_details}\par
\noindent\hspace*{1.5em}\hyperref[app:benchmark_configuration]{\ref*{app:benchmark_configuration}\quad Benchmark Configuration}\dotfill\pageref{app:benchmark_configuration}\par
\noindent\hspace*{1.5em}\hyperref[app:grid_generalization]{\ref*{app:grid_generalization}\quad Grid-Size Generation}\dotfill\pageref{app:grid_generalization}\par
\noindent\hspace*{1.5em}\hyperref[app:geometry_generalization]{\ref*{app:geometry_generalization}\quad Geometry Generalization}\dotfill\pageref{app:geometry_generalization}\par
\noindent\hyperref[app:dynamic_inverse]{\ref*{app:dynamic_inverse}\quad Dynamic Nonlinear Approximate-Inverse Framework}\dotfill\pageref{app:dynamic_inverse}\par
\noindent\hspace*{1.5em}\hyperref[app:inverse_interpretation]{\ref*{app:inverse_interpretation}\quad Approximate-Inverse Interpretation and Transport Structure}\dotfill\pageref{app:inverse_interpretation}\par
\noindent\hspace*{1.5em}\hyperref[app:block_grans_arch]{\ref*{app:block_grans_arch}\quad From GRANS to Block-GRANS}\dotfill\pageref{app:block_grans_arch}\par
\noindent\hyperref[app:supplementary_results]{\ref*{app:supplementary_results}\quad Supplementary Experimental Results}\dotfill\pageref{app:supplementary_results}\par
\noindent\hspace*{1.5em}\hyperref[app:workflow]{\ref*{app:workflow}\quad Preconditioning Integration Experiments}\dotfill\pageref{app:workflow}\par
\noindent\hspace*{1.5em}\hyperref[app:one_basis]{\ref*{app:one_basis}\quad One-Basis Limit Test}\dotfill\pageref{app:one_basis}\par
\bigskip

\section{Uniform Emulation of Fixed-Step GMRES}
\label{app:theory}

This appendix records the theoretical statement: an expressive continuous neural generator can uniformly emulate the correction produced by a fixed number of GMRES steps on a compact, nondegenerate problem family. Namely, a neural basis generator with at least one emitted direction can approximate the residual quality of fixed-step GMRES uniformly over the family.

\paragraph{Theorem 1 (uniform emulation of \(m\)-step GMRES).}
\label{thm:gmres_emulation}
Let \(\mathbb{D}\) be compact and let \(\mathbf{A}(\bm{\mu})\in\mathbb{R}^{n\times n}\) and \(\mathbf{r}_0(\bm{\mu})\in\mathbb{R}^n\) be continuous in \(\bm{\mu}\in\mathbb{D}\). Assume that \(\mathbf{A}(\bm{\mu})\) is uniformly nonsingular and that, for a fixed \(m\), the matrix $\mathbf{Y}_m(\bm{\mu})=\mathbf{A}(\bm{\mu})\mathbf{K}_m(\bm{\mu}),\mathbf{K}_m(\bm{\mu})=[\mathbf{r}_0(\bm{\mu}),\mathbf{A}(\bm{\mu})\mathbf{r}_0(\bm{\mu}),\ldots,\mathbf{A}(\bm{\mu})^{m-1}\mathbf{r}_0(\bm{\mu})]$ has constant rank on \(\mathbb{D}\). Let $\rho_m^{\mathrm{GMRES}}(\bm{\mu})=\min_{\mathbf{y}} \|\mathbf{r}_0(\bm{\mu})-\mathbf{Y}_m(\bm{\mu})\mathbf{y}\|_2=\|\mathbf{r}_0(\bm{\mu})-\mathbf{Y}_m(\bm{\mu})\mathbf{y}_m^*\|_2$, where $\mathbf{y}_m^*=\mathbf{Y}_m(\bm{\mu})^+\mathbf{r}_0$.
Then, for every \(\varepsilon>0\), there exists a continuous neural generator \(\Phi_{\bm{\theta}}\) that maps the system data \((\operatorname{vec}\mathbf{A}(\bm{\mu}),\mathbf{r}_0(\bm{\mu}))\) to a correction vector \(\bm{\phi}_{\bm{\theta}}(\bm{\mu})\in\mathbb{R}^n\) such that the one-dimensional projected update over \(\operatorname{span}\{\bm{\phi}_{\bm{\theta}}(\bm{\mu})\}\) satisfies
$
    \min_{\alpha\in\mathbb{R}}
    \|\mathbf{r}_0(\bm{\mu})-\alpha \mathbf{A}(\bm{\mu})\bm{\phi}_{\bm{\theta}}(\bm{\mu})\|_2
    \le
    \rho_m^{\mathrm{GMRES}}(\bm{\mu})+\varepsilon,
    \forall \bm{\mu}\in\mathbb{D} .
$

\paragraph{Proof.}
Define the compact data set
\[
    \mathbb{S}
    :=
    \{(\operatorname{vec}\mathbf{A}(\bm{\mu}),\mathbf{r}_0(\bm{\mu})):\bm{\mu}\in\mathbb{D}\}
    \subset \mathbb{R}^{n^2+n}.
\]
The set \(\mathbb{S}\) is compact because it is the continuous image of the compact set \(\mathbb{D}\).

First, we show that the GMRES correction is a continuous function of the system data on \(\mathbb{S}\). For any system data \(\mathbf{s}=(\operatorname{vec}\mathbf{A},\mathbf{r})\in\mathbb{S}\), define
\[
    \mathbf{K}_m(\mathbf{A},\mathbf{r}):= [\mathbf{r},\mathbf{A}\mathbf{r},\ldots,\mathbf{A}^{m-1}\mathbf{r}],
    \qquad
    \mathbf{Y}_m(\mathbf{A},\mathbf{r}):= \mathbf{A}\mathbf{K}_m(\mathbf{A},\mathbf{r}).
\]
The entries of \(\mathbf{K}_m(\mathbf{A},\mathbf{r})\) and \(\mathbf{Y}_m(\mathbf{A},\mathbf{r})\) are polynomial functions of the entries of \(\mathbf{A}\) and \(\mathbf{r}\). The rank of \(\mathbf{Y}_m(\mathbf{A},\mathbf{r})\) is constant for \(\mathbf{s}\in\mathbb{S}\), so the Moore--Penrose pseudoinverse \(\mathbf{Y}_m(\mathbf{A},\mathbf{r})^+\) is continuous on \(\mathbb{S}\). Therefore the map
\[
    \mathcal{F}:\mathbb{S}\to\mathbb{R}^n,
    \qquad
    \mathcal{F}(\operatorname{vec}\mathbf{A},\mathbf{r})
    :=
    \mathbf{K}_m(\mathbf{A},\mathbf{r})\mathbf{Y}_m(\mathbf{A},\mathbf{r})^+\mathbf{r}
\]
is continuous. For \(\mathbf{s}(\bm{\mu})=(\operatorname{vec}\mathbf{A}(\bm{\mu}),\mathbf{r}_0(\bm{\mu}))\), this definition gives
\[
    \mathcal{F}(\mathbf{s}(\bm{\mu}))
    =
    \mathbf{K}_m(\bm{\mu})\mathbf{Y}_m(\bm{\mu})^+\mathbf{r}_0(\bm{\mu})
    :=
    \mathbf{z}_m^\star(\bm{\mu}).
\]

Second, we approximate this continuous correction map uniformly. By the universal approximation theorem on the compact set \(\mathbb{S}\), for every \(\delta>0\) there exists a continuous neural network \(\Phi_{\bm{\theta}}:\mathbb{R}^{n^2+n}\to\mathbb{R}^n\) such that
\[
    \sup_{\mathbf{s}\in\mathbb{S}}
    \|\Phi_{\bm{\theta}}(\mathbf{s})-\mathcal{F}(\mathbf{s})\|_2
    <\delta .
\]
For \(\mathbf{s}(\bm{\mu})=(\operatorname{vec}\mathbf{A}(\bm{\mu}),\mathbf{r}_0(\bm{\mu}))\), write
\[
    \bm{\phi}_{\bm{\theta}}(\bm{\mu}):= \Phi_{\bm{\theta}}(\mathbf{s}(\bm{\mu})).
\]
Then
\[
    \sup_{\bm{\mu}\in\mathbb{D}}
    \|\bm{\phi}_{\bm{\theta}}(\bm{\mu})-\mathbf{z}_m^\star(\bm{\mu})\|_2
    <\delta.
\]

Third, we translate correction-vector approximation into residual approximation. Since \(\mathbf{A}\) is continuous and \(\mathbb{D}\) is compact,
\[
    {b_{\max}}:= \sup_{\bm{\mu}\in\mathbb{D}}\|\mathbf{A}(\bm{\mu})\|_2 <\infty .
\]
Uniform nonsingularity implies \(\mathbf{A}(\bm{\mu})\neq 0\) for all \(\bm{\mu}\), so \({b_{\max}}>0\). For each fixed \(\bm{\mu}\), the projected residual over \(\operatorname{span}\{\bm{\phi}_{\bm{\theta}}(\bm{\mu})\}\) is no larger than the residual obtained from the admissible scalar choice \(\alpha=1\). Therefore
\[
\begin{aligned}
    \min_{\alpha\in\mathbb{R}}
    \|\mathbf{r}_0(\bm{\mu})-\alpha \mathbf{A}(\bm{\mu})\bm{\phi}_{\bm{\theta}}(\bm{\mu})\|_2
    &\le
    \|\mathbf{r}_0(\bm{\mu})-\mathbf{A}(\bm{\mu})\bm{\phi}_{\bm{\theta}}(\bm{\mu})\|_2  \\
    &\le
    \|\mathbf{r}_0(\bm{\mu})-\mathbf{A}(\bm{\mu})\mathbf{z}_m^\star(\bm{\mu})\|_2
    +
    \|\mathbf{A}(\bm{\mu})(\mathbf{z}_m^\star(\bm{\mu})-\bm{\phi}_{\bm{\theta}}(\bm{\mu}))\|_2 \\
    &\le
    \rho_m^{\mathrm{GMRES}}(\bm{\mu})
    +
    \|\mathbf{A}(\bm{\mu})\|_2\,
    \|\mathbf{z}_m^\star(\bm{\mu})-\bm{\phi}_{\bm{\theta}}(\bm{\mu})\|_2 \\
    &<
    \rho_m^{\mathrm{GMRES}}(\bm{\mu})+{b_{\max}}\delta .
\end{aligned}
\]
Choose \(\delta=\varepsilon/{b_{\max}}\). Then the last display gives
\[
    \min_{\alpha\in\mathbb{R}}
    \|\mathbf{r}_0(\bm{\mu})-\alpha \mathbf{A}(\bm{\mu})\bm{\phi}_{\bm{\theta}}(\bm{\mu})\|_2
    <
    \rho_m^{\mathrm{GMRES}}(\bm{\mu})+\varepsilon
\]
for every \(\bm{\mu}\in\mathbb{D}\). Replacing the strict inequality by the weaker non-strict inequality gives the stated claim.

The intuition is that fixed-step GMRES defines a continuous correction rule on any compact region where the Krylov action matrix \(\mathbf{Y}_m\) does not change rank. A sufficiently expressive continuous neural generator can uniformly approximate that rule, and the projected one-dimensional update can always choose the scalar \(\alpha=1\), so its residual cannot be worse than the residual of the approximated GMRES correction plus the operator-norm approximation error. \(\square\)

%%%%%%%%%%%%%%%%%%%%%%%%%%%%%%%%%%%%%%%%%%%%%%%%%%%%%%%%%%%%%%%%%%%%%%%%%%%%%%%%%%%%%%%%%%%%%
%%%%%%%%%%%%%%%%%%%%%%%%%%%%%%%%%%%%%%%%%%%%%%%%%%%%%%%%%%%%%%%%%%%%%%%%%%%%%%%%%%%%%%%%%%%%%
%%%%%%%%%%%%%%%%%%%%%%%%%%%%%%%%%%%%%%%%%%%%%%%%%%%%%%%%%%%%%%%%%%%%%%%%%%%%%%%%%%%%%%%%%%%%%
%%%%%%%%%%%%%%%%%%%%%%%%%%%%%%%%%%%%%%%%%%%%%%%%%%%%%%%%%%%%%%%%%%%%%%%%%%%%%%%%%%%%%%%%%%%%%
%%%%%%%%%%%%%%%%%%%%%%%%%%%%%%%%%%%%%%%%%%%%%%%%%%%%%%%%%%%%%%%%%%%%%%%%%%%%%%%%%%%%%%%%%%%%%
\section{Experimental Details}
\label{app:experimental_details}
\subsection{Benchmark Configuration}
\label{app:benchmark_configuration}

All benchmark matrices are generated by a finite-element data-generation pipeline. Each sample is a sparse linear system $\mathbf{A}\mathbf{x}=\mathbf{b}$. The benchmark labels refer to the PDE model used to generate the assembled linear system. All systems are assembled with continuous piecewise-linear triangular finite elements on a quadrilateral domain. The nominal domain has bottom corners $(-1,-1)$ and $(1,-1)$ and top corners $(-1,1)$ and $(1,1)$; for each randomized sample, the two top corners are perturbed independently by a uniform displacement in $[-0.3,0.3]^2$. The mesh also includes a fixed circular interior facet of radius $0.25$ centered at the origin, while the coefficient inclusion used by some PDE branches may be perturbed separately. Nonzero bottom Dirichlet data are handled by a lifting vector \(\mathbf{u}_0\): the stored linear system solved by all iterative methods is the homogeneous-boundary correction system for \(\hat{\mathbf{u}}\), obtained after subtracting \(\mathbf{A}\mathbf{u}_0\) from the assembled right-hand side and enforcing identity rows on boundary degrees of freedom.

For all six tasks, randomized scalar parameters are sampled from their baseline values with relative variation \(\rho=0.3\): each parameter \(\theta_0\) is replaced by \(\theta=\theta_0(1+\xi)\), with \(\xi\sim\mathrm{Unif}[-\rho,\rho]\), unless a smaller geometric perturbation range is stated below. The common diffusion and source pattern is based on a circular inclusion: \[ \kappa(\mathbf{x})=
    \begin{cases}
    \kappa_{\mathrm{in}}, & \|\mathbf{x}-\mathbf{c}\|\le r,\\
    \kappa_{\mathrm{out}}, & \|\mathbf{x}-\mathbf{c}\|>r,
    \end{cases}
    \qquad f(\mathbf{x})=
    \begin{cases}
    s_f, & \|\mathbf{x}-\mathbf{c}\|\le r,\\
    0, & \|\mathbf{x}-\mathbf{c}\|>r.
    \end{cases}
\] Here \(g(x,y)=c_g(1-x^2)\) on the prescribed bottom boundary. For all tasks using this common pattern, \(\kappa_{\mathrm{in}}\in[0.77,1.43]\), \(\kappa_{\mathrm{out}}\in[0.70,1.30]\), \(s_f\in[0.07,0.13]\), and \(c_g\in[0.70,1.30]\).

\begin{table}[H]
\centering
\caption{PDE-derived linear systems used in the main benchmark. Klein--Gordon and wave are implicit one-step systems from second-order time-evolution equations.}
\label{tab:appendix_pde_config}
\footnotesize
\setlength{\tabcolsep}{5pt}
\begin{tabular*}{\linewidth}{@{\extracolsep{\fill}}p{0.24\linewidth}p{0.70\linewidth}@{}}
\toprule
Task & Continuous/discrete form \\
\midrule
Poisson &
\(-\nabla\!\cdot(\kappa\nabla u)=f\). \\
Helmholtz &
\(-\nabla\!\cdot(\kappa\nabla u)+\alpha(\mathbf{x})u=f\). \\
Heat &
Backward Euler step
\((\mathbf{M}+\Delta t \mathbf{K})\mathbf{u}^{n+1}=\mathbf{M}\mathbf{u}^n+\Delta t\,\mathbf{b}\). \\
Advection--diffusion &
\(-\nabla\!\cdot(\kappa\nabla u)+\mathbf{v}\cdot\nabla u=f\). \\
Klein--Gordon &
\((\Delta t^{-2}\mathbf{M}+c^2\mathbf{K}+\mathbf{M}_m)\mathbf{u}^{n+1}=\mathbf{f}^{n+1}+\Delta t^{-2}\mathbf{M}(2\mathbf{u}^n-\mathbf{u}^{n-1})\). \\
Wave &
\((\Delta t^{-2}\mathbf{M}+c^2\mathbf{K})\mathbf{u}^{n+1}=\mathbf{f}^{n+1}+\Delta t^{-2}\mathbf{M}(2\mathbf{u}^n-\mathbf{u}^{n-1})\). \\
\bottomrule
\end{tabular*}
\end{table}

\begin{table}[H]
\centering
\caption{Actual mesh-size distributions for the main benchmark at \(n\approx1000\). Entries are mean\(\pm\)standard deviation with the observed min--max range in parentheses.}
\label{tab:appendix_mesh_stats_n1000}
\small
\setlength{\tabcolsep}{5pt}
\begin{tabular*}{\linewidth}{@{\extracolsep{\fill}}lcc@{}}
\toprule
Task & Train & Test \\
\midrule
Poisson             & $929.35\pm67.81$ (785--1117) & $913.70\pm93.47$ (786--1091) \\
Helmholtz           & $918.72\pm66.86$ (741--1113) & $926.10\pm70.60$ (781--1035) \\
Heat                & $909.88\pm60.63$ (764--1051) & $922.80\pm55.06$ (829--1011) \\
Advection-Diffusion & $913.74\pm55.55$ (761--1042) & $918.00\pm67.28$ (818--1026) \\
Klein-Gordon        & $919.78\pm64.89$ (766--1092) & $935.50\pm50.59$ (892--1076) \\
Wave                & $919.57\pm70.94$ (741--1133) & $941.90\pm60.14$ (842--1020) \\
\bottomrule
\end{tabular*}
\end{table}

% \begin{table}[H]
% \centering
% \caption{Actual mesh-size distributions for the main benchmark at \(n\approx10000\). Entries are mean\(\pm\)standard deviation with the observed min--max range in parentheses.}
% \label{tab:appendix_mesh_stats_n10000}
% \small
% \setlength{\tabcolsep}{5pt}
% \begin{tabular*}{\linewidth}{@{\extracolsep{\fill}}lcc@{}}
% \toprule
% Task & Train & Test \\
% \midrule
% Poisson             & $11139.11\pm779.56$ (9504--12964) & $10794.50\pm1049.34$ (9036--12519) \\
% Helmholtz           & $11178.78\pm802.02$ (9228--13384) & $10707.10\pm658.70$ (9864--11592) \\
% Heat                & $11158.86\pm802.72$ (8886--13347) & $11257.00\pm1044.04$ (9935--13319) \\
% Advection-Diffusion & $11128.69\pm832.55$ (9099--13188) & $11674.40\pm901.93$ (10582--13381) \\
% Klein-Gordon        & $11165.38\pm832.44$ (9002--13012) & $11336.40\pm656.58$ (10243--12460) \\
% Wave                & $11172.76\pm777.30$ (8955--13166) & $11032.60\pm810.63$ (9810--12194) \\
% \bottomrule
% \end{tabular*}
% \end{table}

\noindent\textbf{Poisson.} The Poisson task uses the common diffusion/source inclusion and boundary lifting described above. Its only task-specific geometric perturbations are the coefficient-inclusion radius \(r=0.25(1+\xi_r)\), where \(\xi_r\sim\mathrm{Unif}[-0.15,0.15]\), and the two coordinates of the inclusion center, each sampled in \([-0.03,0.03]\). The assembled correction system is the P1 finite-element stiffness system \(\mathbf{K}\hat{\mathbf{u}}=\tilde{\mathbf{b}}\), where local stiffness entries are obtained from \(\int_{\mathbb{T}} \kappa\nabla\phi_i\cdot\nabla\phi_j\) and the load uses the centroid value of the source on each triangle.

\noindent\textbf{Helmholtz.} The Helmholtz task uses the same diffusion/source inclusion pattern as Poisson but adds a positive reaction term. The reaction parameters are sampled as \(\alpha_{\mathrm{in}}\in[1.40,2.60]\) and \(\alpha_{\mathrm{out}}\in[0.70,1.30]\). The coefficient-inclusion radius again uses \(r=0.25(1+\xi_r)\), \(\xi_r\sim\mathrm{Unif}[-0.15,0.15]\), while each coordinate of the inclusion center is sampled in \([-0.06,0.06]\). The discrete correction matrix is \(\mathbf{K}+\mathbf{M}_{\alpha}\), where \(\mathbf{M}_{\alpha}\) is the consistent P1 weighted mass matrix assembled with centroid-valued \(\alpha(\mathbf{x})\). This benchmark task is therefore a positive-reaction Helmholtz-form sparse linear system, not the indefinite high-frequency Helmholtz regime.

\noindent\textbf{Heat.} The heat task represents one backward-Euler implicit step, not a long time-dependent simulation. It uses the common diffusion/source inclusion with the nominal center \(\mathbf{c}=(0,0)\) and radius \(0.25\). The time step satisfies \(\Delta t\in[0.07,0.13]\). The previous state is parameterized as
\[
u^n(x,y)=a_1\sin(\pi x)\sin(\pi y)+a_2\cos(\pi x)\cos(\pi y),
\]
with \(a_1\in[0.70,1.30]\) and \(a_2\in[0.35,0.65]\), and is made consistent with the prescribed bottom boundary before forming the step. The resulting correction system uses \(\mathbf{A}=\mathbf{M}+\Delta t \mathbf{K}\) and right-hand side \(\mathbf{M}\mathbf{u}^n+\Delta t \mathbf{b}\), followed by the same lifting and boundary-row enforcement as the other finite-element tasks.

\noindent\textbf{Advection--Diffusion.} The advection--diffusion task uses the common diffusion/source inclusion with nominal center \(\mathbf{c}=(0,0)\) and radius \(0.25\), and adds a first-order transport field \(\mathbf{v}(x,y)=\beta(-y,x)\), where \(\beta\in[0.35,0.65]\). The discrete correction matrix is \(\mathbf{K}+\mathbf{C}\). The diffusion block is the same stiffness matrix used in the Poisson branch, while the nonsymmetric transport block is assembled from the centroid quadrature approximation to \(\int_{\mathbb{T}} \phi_i\,\mathbf{v}\cdot\nabla\phi_j\). This is the only main benchmark branch whose nonsymmetry comes from a first-order differential operator.

\noindent\textbf{Klein--Gordon.} The Klein--Gordon task represents one implicit step of the second-order equation \(u_{tt}-c^2\nabla\!\cdot(\kappa\nabla u)+m^2(\mathbf{x})u=f\). It uses the common diffusion/source inclusion with nominal center \(\mathbf{c}=(0,0)\) and radius \(0.25\), wave speed \(c\in[0.70,1.30]\), time step \(\Delta t\in[0.35,0.65]\), and mass coefficients \(m^2_{\mathrm{in}}\in[2.10,3.90]\), \(m^2_{\mathrm{out}}\in[1.05,1.95]\). Two smooth previous states \(u^n\) and \(u^{n-1}\) are sampled from sinusoidal modes. The assembled correction matrix is \(\Delta t^{-2}\mathbf{M}+c^2\mathbf{K}+\mathbf{M}_m\), and the right-hand side contains \(\mathbf{f}^{n+1}+\Delta t^{-2}\mathbf{M}(2\mathbf{u}^n-\mathbf{u}^{n-1})\).

\noindent\textbf{Wave.} The wave task represents one implicit step of \(u_{tt}-c^2\nabla\!\cdot(\kappa\nabla u)=f\). It uses the common diffusion/source inclusion, wave speed \(c\in[0.70,1.30]\), time step \(\Delta t\in[0.35,0.65]\), and two smooth previous states sampled from the same sinusoidal family used in the Klein--Gordon task. The assembled correction matrix is \(\Delta t^{-2}\mathbf{M}+c^2\mathbf{K}\), and the right-hand side is \(\mathbf{f}^{n+1}+\Delta t^{-2}\mathbf{M}(2\mathbf{u}^n-\mathbf{u}^{n-1})\).

For each PDE and each scale, the training set contains 200 samples and the held-out inference set contains 10 samples. The nominal mesh scales are \(n\approx1000\), but the actual number of nodes varies because both the domain geometry and the triangular refinement are randomized. Tables~\ref{tab:appendix_mesh_stats_n1000} report the realized distributions recorded by the data-generation logs.

\noindent\textbf{Model and training parameters.} The training configuration in the benchmark script uses the same schedule for RINS and GRANS: 100 stage-1 epochs, 100 stage-2 epochs, and 150 stage-3 epochs, with learning rate \(10^{-3}\), and weight decay \(10^{-5}\). The main RINS model uses 20 generated filters per neural iteration, polynomial order \(L=4\), Chebyshev initialization, gate hidden width 16, and gate scale 0.5. Static-RINS uses the same polynomial generator without the pointwise gate. GRANS uses coordinate dimension 2, positional encoding dimension 16, 48 hidden channels, 4 graph layers, 5 basis directions per layer, 3 attention heads, and dropout 0.08. During inference, both neural solvers are run for 200 neural block iterations in the main benchmark. 

\noindent\textbf{Inference and statistic setting.} For timing, the benchmark fixes CPU linear-algebra threading with \texttt{OMP\_NUM\_THREADS=1} and \texttt{MKL\_NUM\_THREADS=1}; GPU timings are recorded with explicit CUDA synchronization around the measured region and wall-clock intervals from \texttt{time.perf\_counter()}. These synchronization barriers prevent unfinished CUDA kernels from being hidden behind asynchronous host execution. An optimized deployment that pipelines these operations asynchronously would typically reduce overhead or improve throughput, so the reported neural-solver times should be viewed as conservative synchronized measurements. The paper figures report median curves over the 10 held-out test samples for a fixed training seed, with IQR bands over those samples. The timing tables pool all valid seed/sample time-to-threshold observations: for each available seed and each held-out sample, we record one value and compute the reported mean and standard deviation over the pooled observations. Thus, the reported standard deviation includes variation across seeds and variation across samples under each seed. Timed GPU sections call \path{torch.cuda.synchronize()} before and after the measured solver step.

%%%%%%%%%%%%%%%%%%%%%%%%%%%%%%%%%%%%%%%%%%%%%%%%%%%%%%%%%%%%%%%%%%%%%%%%%%%%%%%%%%%%%%%%%%%%%
%%%%%%%%%%%%%%%%%%%%%%%%%%%%%%%%%%%%%%%%%%%%%%%%%%%%%%%%%%%%%%%%%%%%%%%%%%%%%%%%%%%%%%%%%%%%%
%%%%%%%%%%%%%%%%%%%%%%%%%%%%%%%%%%%%%%%%%%%%%%%%%%%%%%%%%%%%%%%%%%%%%%%%%%%%%%%%%%%%%%%%%%%%%
%%%%%%%%%%%%%%%%%%%%%%%%%%%%%%%%%%%%%%%%%%%%%%%%%%%%%%%%%%%%%%%%%%%%%%%%%%%%%%%%%%%%%%%%%%%%%
%%%%%%%%%%%%%%%%%%%%%%%%%%%%%%%%%%%%%%%%%%%%%%%%%%%%%%%%%%%%%%%%%%%%%%%%%%%%%%%%%%%%%%%%%%%%%

\subsection{Grid-Size Generation}
\label{app:grid_generalization}

\begin{figure}[!htbp]
    \centering
    \includegraphics[width=\linewidth]{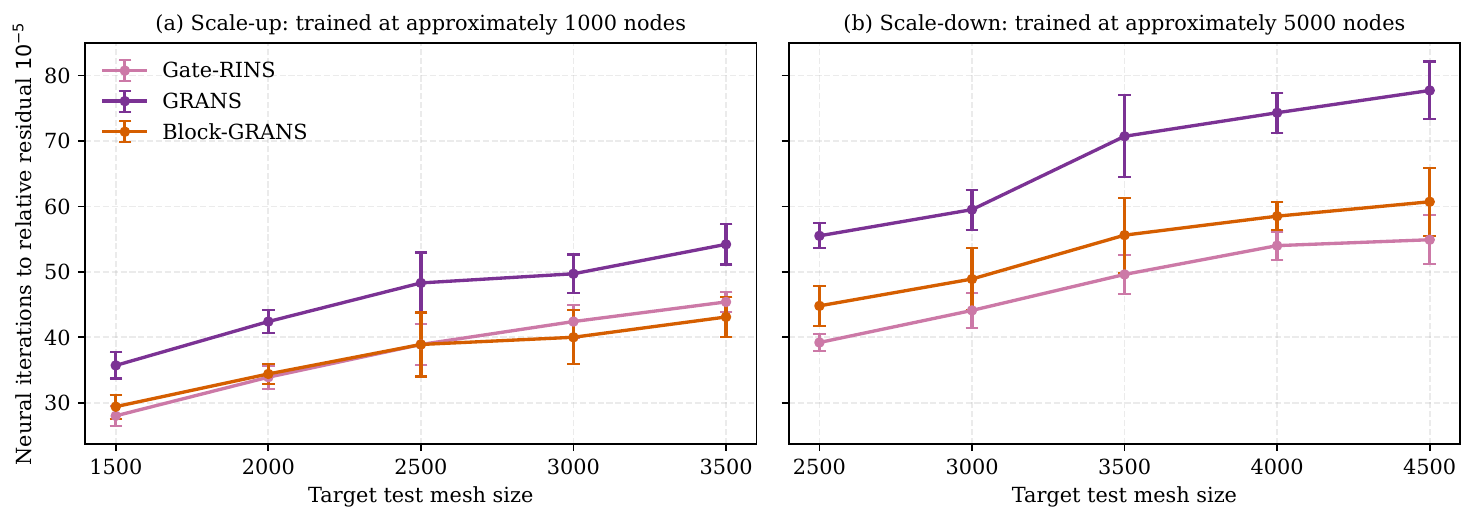}
    \caption{Absolute iteration counts for Poisson grid-size generalization: (a) training at approximately 1000 nodes and testing on larger meshes; (b) training at approximately 5000 nodes and testing on smaller meshes. Points and error bars denote means and sample standard deviations over 10 test systems.}
    \label{fig:grid_generalization_absolute}
\end{figure}

Figure~\ref{fig:grid_generalization_absolute} complements the baseline-relative curves in Section~\ref{sec:ood_generalization} with absolute neural iteration counts to relative residual $10^{-5}$. The two experiments use separate controllers trained at approximately 1000 and 5000 nodes, respectively, with a fixed training seed. Each plotted mean and sample standard deviation is computed over 10 test systems at the indicated target mesh size. 

\paragraph{Transfer to larger systems.} The training-scale mean counts are 22.5, 28.3, and 23.6 for Gate-RINS, GRANS, and Block-GRANS, respectively. At target size 3500, these become $45.4\pm1.5$, $54.2\pm3.1$, and $43.1\pm3.0$. Gate-RINS consistently uses fewer iterations than GRANS, but its ordering relative to Block-GRANS is not invariant: the two tie at 2500 nodes (38.9 iterations), and Block-GRANS has lower mean counts at 3000 and 3500 nodes.

\paragraph{Transfer to smaller systems.} Training-scale means are 60.5, 83.3, and 61.8 in the same model order. At target size 2500, the corresponding counts decrease to $39.2\pm1.3$, $55.5\pm1.9$, and $44.8\pm3.1$. Gate-RINS has the lowest mean count at every tested size in this experiment. 

\subsection{Geometry Generalization}
\label{app:geometry_generalization}

This appendix records the detailed protocol behind Panel~(c) of Figure~\ref{fig:exp_generation} (and its summary in Section~\ref{sec:ood_generalization}). The experiment isolates whether the learned controllers keep using spatial information, rather than memorizing a particular domain, by holding the PDE family fixed and changing the mesh geometry.

\paragraph{Domain geometry.} Training systems are assembled on a rectangular domain with corners (-1,-1), (1,-1), (1,1), (-1,1). Test systems are assembled on an I-shaped domain whose boundary is the polygon (-1,-1), (1,-1), (1,-0.3), (0.2,-0.3), (0.2,0.3), (1,0.3), (1,1), (-1,1), (-1,0.3), (-0.2,0.3), (-0.2,-0.3), (-1,-0.3). Both domains are triangulated with meshpy.triangle to roughly \(n\approx1000\) nodes using an area-based refinement criterion proportional to \(250/n_{\mathrm{target}}\), so the two geometries yield comparable discretization resolution.

\paragraph{Equation and randomized parameters.} All samples solve the heterogeneous Poisson equation
\[
-\nabla\cdot\bigl(\kappa(\mathbf{x})\nabla u\bigr)=f(\mathbf{x})\quad\text{in }\Omega,
\qquad
u(\mathbf{x})=g(\mathbf{x})\quad\text{on }y=-1,
\]
with the same circular inclusion of radius \(0.25\) centered at the origin. Inside the inclusion \(\kappa=\kappa_{\mathrm{inner}}\) and the source \(f=s\); outside, \(\kappa=\kappa_{\mathrm{outer}}\) and \(f=0\), with bottom Dirichlet data \(g=\beta(1-x^2)\). For each training and test sample, the two diffusion coefficients, the source strength, and the boundary scale are each drawn as $\theta=\theta_{\mathrm{base}}\cdot\bigl(1+\delta\,\mathcal{U}(-1,1)\bigr), \delta=0.3$ with baselines \(\kappa_{\mathrm{inner}}=1.1\), \(\kappa_{\mathrm{outer}}=1.0\), \(s=0.1\), and \(\beta=1.0\). Thus both the rectangular training set and the I-shaped test set carry the same \(\pm30\%\) parameter variation. Only the mesh geometry is shifted between them.

\paragraph{Discretization and linear system.} Continuous piecewise-linear triangular finite elements assemble the stiffness matrix \(\mathbf{A}\) and the body-force vector \(b\). Nonzero bottom Dirichlet data are lifted by a solution \(u_0\) supported on the boundary nodes (those with \(y=-1\)), and the iterative solvers act on the homogeneous correction system for \(\widehat{u}=u-u_0\), after forcing identity rows on all boundary degrees of freedom. Each stored sample records the COO form of \(\mathbf{A}\), the right-hand side \(\mathbf{b}\), the node coordinates, and the direct solve \(u\). The training set contains 200 samples on the rectangle, and the held-out inference set contains 20 samples on the I-shape; a single training seed \(2026\) is used throughout.

\paragraph{Training.} Gate-RINS, GRANS, and Block-GRANS are trained on the 200 rectangular training systems with the same three-stage schedule as the main benchmark: 100 stage-1 epochs, 100 stage-2 epochs, and 150 stage-3 epochs, with AdamW, learning rate \(10^{-3}\), weight decay \(10^{-5}\), and gradient clipping at norm 1.0. Gate-RINS uses 20 filters, Chebyshev order \(4\), coordinate dimension 2, gate hidden width 16, and gate scale 0.5. GRANS uses coordinate dimension 2, positional encoding dimension 16, 48 hidden channels, 4 graph layers, 5 basis directions per layer, 3 attention heads, and dropout 0.08; Block-GRANS adds a structured feature-block message projection with 4 projection blocks, matching the configuration in Section~\ref{app:block_grans_arch}.

\paragraph{Inference and figure.} Each model is evaluated on the same 20 I-shaped test systems, generating 20 basis directions per neural iteration for a budget of 50 iterations with a maximum accumulated basis size of 1000. Figure~\ref{fig:exp_generation}c reports the per-iteration relative residual averaged over the 20 samples, with the shaded band spanning the 25th--75th percentile; panels (a) and (b) of the same figure use the shared method-color legend described in the caption. Per-model residual histories are cached as \texttt{npz} files and reused unless inference is explicitly re-run.

\section{Dynamic Nonlinear Approximate-Inverse Framework}
\label{app:dynamic_inverse}
\subsection{Approximate-Inverse Interpretation and Transport Structure}
\label{app:inverse_interpretation}

\paragraph{What is approximated.} Consider a nonsingular linear system
$\mathbf{A}\mathbf{x}=\mathbf{b}$, and let $\mathbf{r}_t$ denote the residual
at iteration $t$. The ideal correction to the current iterate is
$\mathbf{A}^{-1}\mathbf{r}_t$. Since $\mathbf{A}^{-1}$ is precisely the object
the iteration aims to approximate---a target rather than an available
input---the neural solver cannot invoke it directly. Instead, it generates a
raw block $\mathbf{B}_t$ and selects the minimum-residual correction within
$\mathrm{span}(\mathbf{B}_t)$, computed by least squares:
% \begin{equation}
$
    \mathbf{d}_t=\mathbf{B}_t(\mathbf{A}\mathbf{B}_t)^+\mathbf{r}_t,
    % \qquad
    \|\mathbf{r}_t-\mathbf{A}\mathbf{d}_t\|_2
    =\min_{\mathbf{y}}\|\mathbf{r}_t-\mathbf{A}\mathbf{B}_t\mathbf{y}\|_2.
$
% \end{equation}
The inverse approximation is thus selected by projection onto the generated
block, rather than prescribed independently for each column. 

\paragraph{Gate-RINS: diagonal gating of local polynomial actions.}
Here we follow the description in the Methodology while omitting the subscript $t$ that denotes the iteration step. For Gate-RINS generating $m$ basis vectors at each neural iteration, the block is $\mathbf{B}=[\mathbf{D}_1\mathbf{b}_1,\cdots,\mathbf{D}_m\mathbf{b}_m]$, where $\mathbf{D}_j$ is a diagonal matrix, $\mathbf{D}_j=\operatorname{diag}(g_{1j},\cdots,g_{nj})$, and $\mathbf{b}_j=\sum_{\ell=0}^{l}\alpha_{j\ell}\mathbf{c}^{(\ell)}$. By absorbing the coefficients of the Chebyshev polynomials into $\alpha$, we obtain $\mathbf{b}_j=\sum_{\ell=0}^{l}\alpha_{j\ell}\mathbf{A}^{\ell}\mathbf{r}$. Each column of the block $\mathbf{B}$ can therefore be written as $\sum_{\ell=0}^{l}\alpha_{j\ell}\mathbf{D}_j\mathbf{A}^{\ell}\mathbf{r}$. If we restrict the model to generating one basis vector per iteration, omit the $\ell=0$ term, and set $l=1$, the resulting basis reduces to $\mathbf{D}\mathbf{A}\mathbf{r}$, which can be interpreted as an inverse-like residual-adapted action related to preconditioned Krylov iteration. When $\mathbf{D}$ commutes with $\mathbf{A}$, this can further be written as $\mathbf{D}\mathbf{A}=\mathbf{A}\mathbf{D}=\mathbf{A}\mathbf{M}^{-1}$ by identifying $\mathbf{M}^{-1}=\mathbf{D}$. The distinction is that $\mathbf{D}$ is residual-adaptive: $\mathbf{D}=\mathbf{D}(\mathbf{r}_t)$.

\paragraph{GRANS: sparse propagation in a lifted feature space.}
GRANS exhibits richer structural patterns. \textbf{Firstly}, we restrict $l=1$, corresponding to one message-passing layer per neural iteration, and omit the subscript $t$ denoting the iteration step. $\mathbf{B}=\operatorname{MLP}(\mathbf{h})\in\mathbb{R}^{n\times q}$, where $\mathbf{h}_i^{l+1}=\sum_{j\in\mathcal{N}_i}\gamma_{ij}\mathbf{W}\mathbf{h}_j^l\in\mathbb{R}^q$, with $i,j=1,\cdots,n$. \textbf{Secondly}, let $q=1$, namely, the embedding dimension of each node is reduced to one. Then $\mathbf{h}_i$ degenerates from a vector to a scalar, and the matrix $\mathbf{W}$ becomes a scalar as well. We then have
\begin{equation}
    h_i^{l+1}=\sum_{j\in\mathcal{N}_i}\gamma_{ij}h_j^l,
    \label{scalar_propagation}
\end{equation}
where $W$ has been absorbed into $\gamma$. \textbf{Thirdly}, ignoring the coordinate input $\mathbf{X}_{\mathrm{coord}}$, we have $h_i=\beta r_i$, $i=1,\cdots,n$, where $\beta$ is a scalar induced by the MLP and normalization. We then have $h_i^{l+1}=\sum_{j\in\mathcal{N}_i}\gamma_{ij}h_j^l$, namely, $\mathbf{h}^{l+1}=\mathbf{\Gamma}\mathbf{h}^l\in\mathbb{R}^n$, where $\mathbf{\Gamma}$ shares the same sparsity pattern as $\mathbf{A}$. Under the restriction $l=1$, this gives $\mathbf{h}=\mathbf{\Gamma}\mathbf{r}$. Again, this can be interpreted as a residual-dependent preconditioning action, where $\mathbf{\Gamma}=\mathbf{\Gamma}_{\theta}(\mathbf{A},\mathbf{r})$ is trained to approximate the action of $\mathbf{A}^{-1}$. The distinction is that $\mathbf{\Gamma}$ is residual-adaptive. For multilayer propagation, the output takes the form $\mathbf{h}=\mathbf{\Gamma}_1\mathbf{\Gamma}_2\cdots\mathbf{\Gamma}_l\mathbf{r}$, whose effective sparsity pattern can become denser than that of $\mathbf{A}$.

\textbf{Lifting to the tensor-product space.}
When $q=1$, $\mathbf{\Gamma}\in\mathbb{R}^{n\times n}$ is a sparse matrix sharing the same sparsity pattern as $\mathbf{A}$. When $q>1$, the corresponding propagation operator acts in $\mathbb{R}^{nq}$ and can be represented as a block-sparse matrix in $\mathbb{R}^{nq\times nq}$. Each node is associated with a $q$-dimensional embedding $\mathbf{h}_i\in\mathbb{R}^q$. The original node space $\mathbb{R}^n$ is lifted to the tensor-product space $\mathbb{R}^n\otimes\mathbb{R}^q$, where the first factor indexes the nodes and the second represents the latent feature space within each node. High-dimensional GRANS can therefore be interpreted as learning a block-sparse operator in a lifted state space, rather than directly learning an ordinary sparse matrix in the original space $\mathbb{R}^n$. Given an original residual $\mathbf{r}\in\mathbb{R}^n$, it is first lifted through an encoding map $E$ to $\mathbf{H}\in\mathbb{R}^{n\times q}$, then propagated in the lifted space by an operator $\mathcal{D}_{\theta}$, and finally projected back to the physical space through a projection head $\Pi$. The overall process can therefore be written as
\[
\mathbf{r}\xrightarrow{E}\mathbf{H}\xrightarrow{\mathcal{D}_{\theta}}\mathbf{H}'\xrightarrow{\Pi}\mathbf{q},
\]
so that the effective operator acting on the original physical space can be written as
\[
D_{\mathrm{eff}}(\mathbf{r})=\Pi\mathcal{D}_{\theta}(\mathbf{r})E.
\]
From this perspective, the high-dimensional embedding can be viewed as a lifting mechanism: the original problem is first lifted to a higher-dimensional space endowed with auxiliary internal degrees of freedom, block-sparse dynamics are then performed in the lifted space, and the result is finally projected back to the original solution space.

\textbf{The limits of commutativity.}
For $q=1$, the coefficients $\gamma_{ij}$ in (Eq.~\eqref{scalar_propagation}) are scalars and therefore naturally satisfy $\gamma_{ij}\gamma_{ji}=\gamma_{ji}\gamma_{ij}$. This property does not generally hold after lifting to the tensor-product space, where $\mathbf{\Gamma}_{ij}\mathbf{\Gamma}_{ji}-\mathbf{\Gamma}_{ji}\mathbf{\Gamma}_{ij}\triangleq[\mathbf{\Gamma}_{ij},\mathbf{\Gamma}_{ji}]\neq0$. We refer to this property as the \emph{path-propagation noncommutativity} of GRANS. We acknowledge that its theoretical role is not yet fully understood. We leave a systematic theoretical investigation of this property to future work. Motivated by this observation, we design a variant of GRANS, termed Block-GRANS, to partially restrict this path-propagation noncommutativity. As shown in this paper, Block-GRANS exhibits distinct empirical properties and achieves better performance in most of the evaluated settings.

\subsection{From GRANS to Block-GRANS}
\label{app:block_grans_arch}

At neural iteration $t$, the model receives the current residual $\mathbf{r}_t$, the sparse matrix graph induced by $\mathbf{A}$, edge attributes $a_{ij}$, and node coordinates, and returns a multi-column basis block that is inserted into the same projected minimum-residual closure. Block-GRANS modifies only the admissible graph basis generator while leaving the algebraic update unchanged.

\textbf{Restating GRANS.} In GRANS, node coordinates are first mapped to a learnable positional encoding $\mathbf{z}_i=\phi_{\mathrm{PE}}(\mathbf{x}_i)$, and the initial node feature is constructed as $\mathbf{h}_i^{(0)}=[\widetilde r_{t,i},\mathbf{z}_i]$, where $\widetilde{\mathbf{r}}_t$ denotes the per-graph normalized residual. For each graph edge $(j,i)$ induced by a nonzero entry of $\mathbf{A}$, a GATv2 layer computes an attention weight from the target feature, source feature, and edge value $a_{ij}$. Abstractly, one can write
\begin{equation}
    \gamma_{ij}^{(\ell)}
    =
    \operatorname{softmax}_{j\in\mathcal{N}(i)}
    a_{\bm{\theta}}^{(\ell)}
    \bigl(
    \mathbf{h}_i^{(\ell-1)},
    \mathbf{h}_j^{(\ell-1)},
    a_{ij}
    \bigr),
    \qquad
    \widehat{\mathbf{h}}_i^{(\ell)}
    =
    \sum_{j\in\mathcal{N}(i)}
    \alpha_{ij}^{(\ell)}
    \mathbf{W}_s^{(\ell)}
    \mathbf{h}_j^{(\ell-1)} .
\end{equation}
After batch normalization and an ELU nonlinearity, each layer has its own projection head $\psi_\ell$, producing $\mathbf{V}^{(\ell)}=\psi_\ell(\mathbf{H}^{(\ell)})\in\mathbb{R}^{n\times q}$. The raw GRANS block is then formed by concatenating the outputs from all message-passing layers,
% \begin{equation}
$
    \mathbf{V}_{\mathrm{raw}}
    =
    [\mathbf{V}^{(1)},\mathbf{V}^{(2)},\ldots,\mathbf{V}^{(l)}]
    \in\mathbb{R}^{n\times lq},
% \end{equation}
$
followed by graph-wise QR orthogonalization before the least-squares projection step. In the default setting, $l=4$ and $q=5$, so one forward pass emits $20$ basis directions.

\textbf{Characterizing Block-GRANS.} Block-GRANS preserves the same input construction, edge attention, per-layer basis heads, concatenation, and QR orthogonalization as GRANS. The only architectural modification is imposed on the source feature transformation inside each attention layer. Instead of using a dense source projection $\mathbf{W}_s^{(\ell)}$, Block-GRANS partitions the latent feature channels into $p_{\mathrm{blk}}$ contiguous blocks and replaces the dense projection by a block-diagonal map,
\begin{equation}
    \mathbf{W}_{\mathrm{blk}}^{(\ell)}
    =
    \operatorname{diag}
    \bigl(
    \mathbf{W}_1^{(\ell)},
    \mathbf{W}_2^{(\ell)},
    \ldots,
    \mathbf{W}_{p_{\mathrm{blk}}}^{(\ell)}
    \bigr).
\end{equation}
If the node embedding follows the same partition,
\(
\mathbf{h}_j^{(\ell-1)}
=
[
\mathbf{h}_{j,1}^{(\ell-1)},
\ldots,
\mathbf{h}_{j,p_{\mathrm{blk}}}^{(\ell-1)}
],
\)
then the propagated message becomes
% \begin{equation}
$
    \mathbf{m}_{ij}^{(\ell)}
    =
    \alpha_{ij}^{(\ell)}
    \mathbf{W}_{\mathrm{blk}}^{(\ell)}
    \mathbf{h}_j^{(\ell-1)}
    =
    \alpha_{ij}^{(\ell)}
    \bigl[
    \mathbf{W}_1^{(\ell)}
    \mathbf{h}_{j,1}^{(\ell-1)},
    \ldots,
    \mathbf{W}_{p_{\mathrm{blk}}}^{(\ell)}
    \mathbf{h}_{j,p_{\mathrm{blk}}}^{(\ell-1)}
    \bigr].
% \end{equation}
$
The layer then aggregates these messages over the same graph induced by $\mathbf{A}$,
% \begin{equation}
$
    \widehat{\mathbf{h}}_i^{(\ell)}
    =
    \sum_{j\in\mathcal{N}(i)}
    \mathbf{m}_{ij}^{(\ell)},
% \end{equation}
$
while the remaining forward computation is identical to GRANS:
% \begin{equation}
$
    \mathbf{B}_t^{\mathrm{Block\text{-}GRANS}}
    =
    \operatorname{qr}
    \left(
    [
    \psi_1(\mathbf{H}^{(1)}),
    \ldots,
    \psi_l(\mathbf{H}^{(l)})
    ]
    \right).
% \end{equation}
$

From the lifted-space perspective introduced above, standard GRANS applies dense transformations within the latent feature space $\mathbb{R}^q$ at each graph propagation step, allowing arbitrary cross-channel mixing before information is transported to neighboring nodes. Block-GRANS restricts this internal transformation to a direct sum of lower-dimensional subspaces,
\[
\mathbb{R}^q
=
\mathcal{H}_1\oplus\cdots\oplus\mathcal{H}_{p_{\mathrm{blk}}},
\]
so that each source transformation acts independently within the corresponding latent block. Equivalently, the block-sparse operator in the lifted space retains the same spatial graph connectivity as GRANS, while its internal channel transport is restricted by the block-diagonal structure.

\textbf{The tunable path-propagation commutativity of Block-GRANS.} With $p_{\mathrm{blk}}=1$, $\mathbf{W}_{\mathrm{blk}}^{(\ell)}$ reduces to the original dense source projection and Block-GRANS recovers the standard GRANS source transformation. As $p_{\mathrm{blk}}$ increases, cross-block channel mixing is progressively removed, reducing the degrees of freedom of the latent transport operator. In the limiting case where each block contains a single latent channel, the internal source transformation becomes strictly diagonal. Absorbing the scalar attention coefficient $\gamma_{ij}$ into the diagonal map yields a local transport operator $\mathbf{\Gamma}_{ij}=\gamma_{ij}\mathbf{W}_{ij}$, which is also diagonal. Therefore, for any two local transport operators,
\[
[\mathbf{\Gamma}_{ij},\mathbf{\Gamma}_{kl}]=0,
\]
and the local path-propagation commutativity in the latent feature space is recovered.

\section{Supplementary Experimental Results}
\label{app:supplementary_results}
\subsection{Preconditioning Integration Experiments}
\label{app:workflow}

To test whether RINS composes with an existing numerical acceleration technique, we repeat the six-PDE \(n\approx1000\) benchmark after applying the same left Jacobi preconditioner to every training and inference linear system. The preconditioned datasets are deterministic transforms of the main benchmark splits, so all three methods see the same matrices and right-hand sides after preconditioning. Gate-RINS and GRANS are trained on the 200 preconditioned training systems for each PDE with the same \(50/50/100\) schedule and the same 20-direction-per-iteration budget. At inference time, preconditioned RINS, preconditioned GRANS, and preconditioned GMRES are evaluated on the same 10 held-out systems. As in the main benchmark, GMRES is stopped when its relative residual reaches the better final residual obtained by RINS and GRANS on that PDE, or when the maximum iteration budget is exhausted.

\begin{figure}[!htbp]
    \centering
    \IfFileExists{figures/main_six_pdes_n1000_left_jacobi.pdf}{%
        \includegraphics[width=\linewidth]{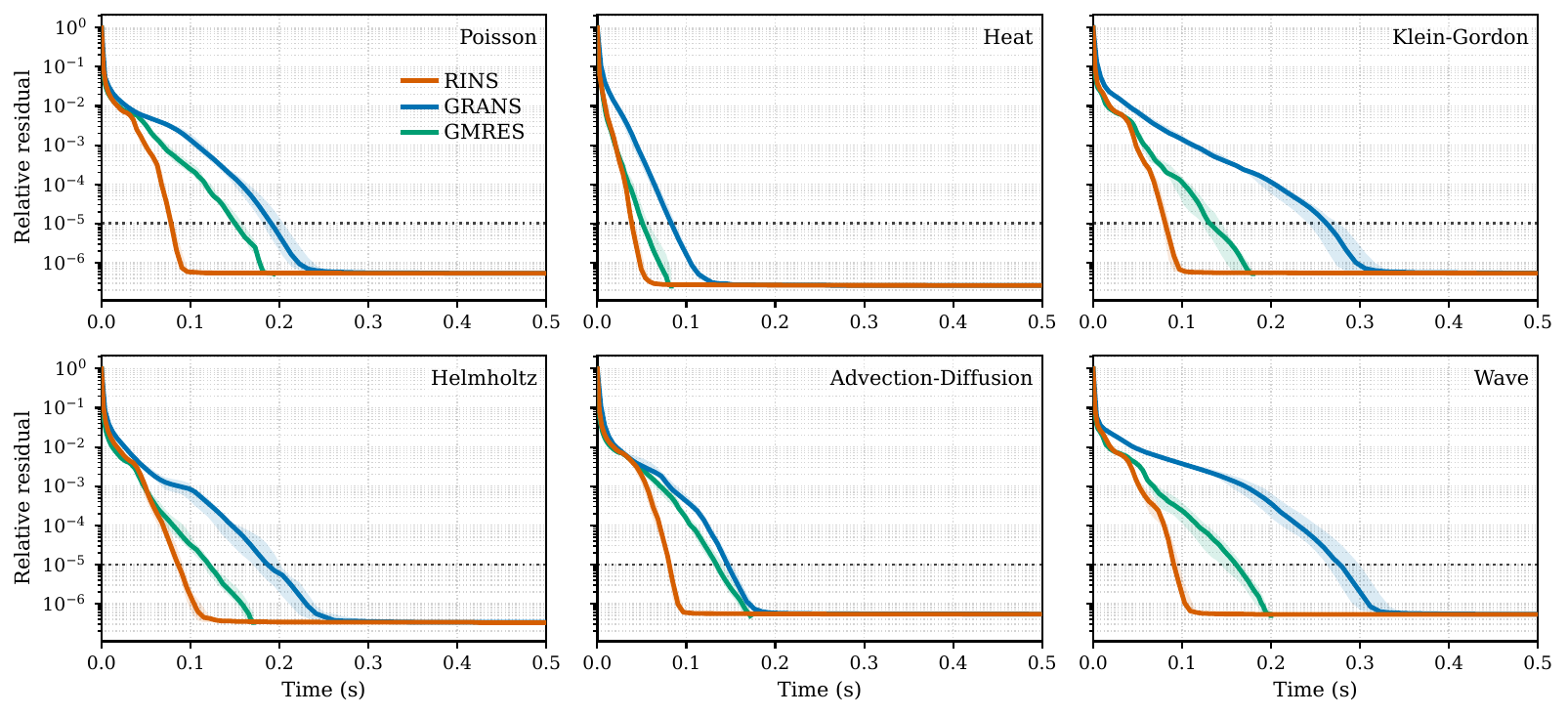}%
    }{%
        \fbox{\parbox[c][0.20\textheight][c]{0.82\linewidth}{\centering Preconditioned six-PDE residual-time figure generated on the server.}}%
    }
    \caption{Preconditioned six-PDE inner-solve workflow at \(n\approx1000\). RINS, GRANS, and GMRES are all evaluated on the same left-Jacobi-preconditioned systems; curves report relative residual versus cumulative inner linear-solve time.}
    \label{fig:preconditioned_poisson}
\end{figure}

%%%%%%%%%%%%%%%%%%%%%%%%%%%%%%%%%%%%%%%%%%%%%%%%%%%%%%%%%%%%%%%%%%%%%%%%%%%%%%%%%%%%%%%%%%%%%
%%%%%%%%%%%%%%%%%%%%%%%%%%%%%%%%%%%%%%%%%%%%%%%%%%%%%%%%%%%%%%%%%%%%%%%%%%%%%%%%%%%%%%%%%%%%%
%%%%%%%%%%%%%%%%%%%%%%%%%%%%%%%%%%%%%%%%%%%%%%%%%%%%%%%%%%%%%%%%%%%%%%%%%%%%%%%%%%%%%%%%%%%%%
%%%%%%%%%%%%%%%%%%%%%%%%%%%%%%%%%%%%%%%%%%%%%%%%%%%%%%%%%%%%%%%%%%%%%%%%%%%%%%%%%%%%%%%%%%%%%
%%%%%%%%%%%%%%%%%%%%%%%%%%%%%%%%%%%%%%%%%%%%%%%%%%%%%%%%%%%%%%%%%%%%%%%%%%%%%%%%%%%%%%%%%%%%%
\subsection{One-Basis Limit Test}
\label{app:one_basis}

\begin{figure}[!ht]
    \centering
    \includegraphics[width=\textwidth]{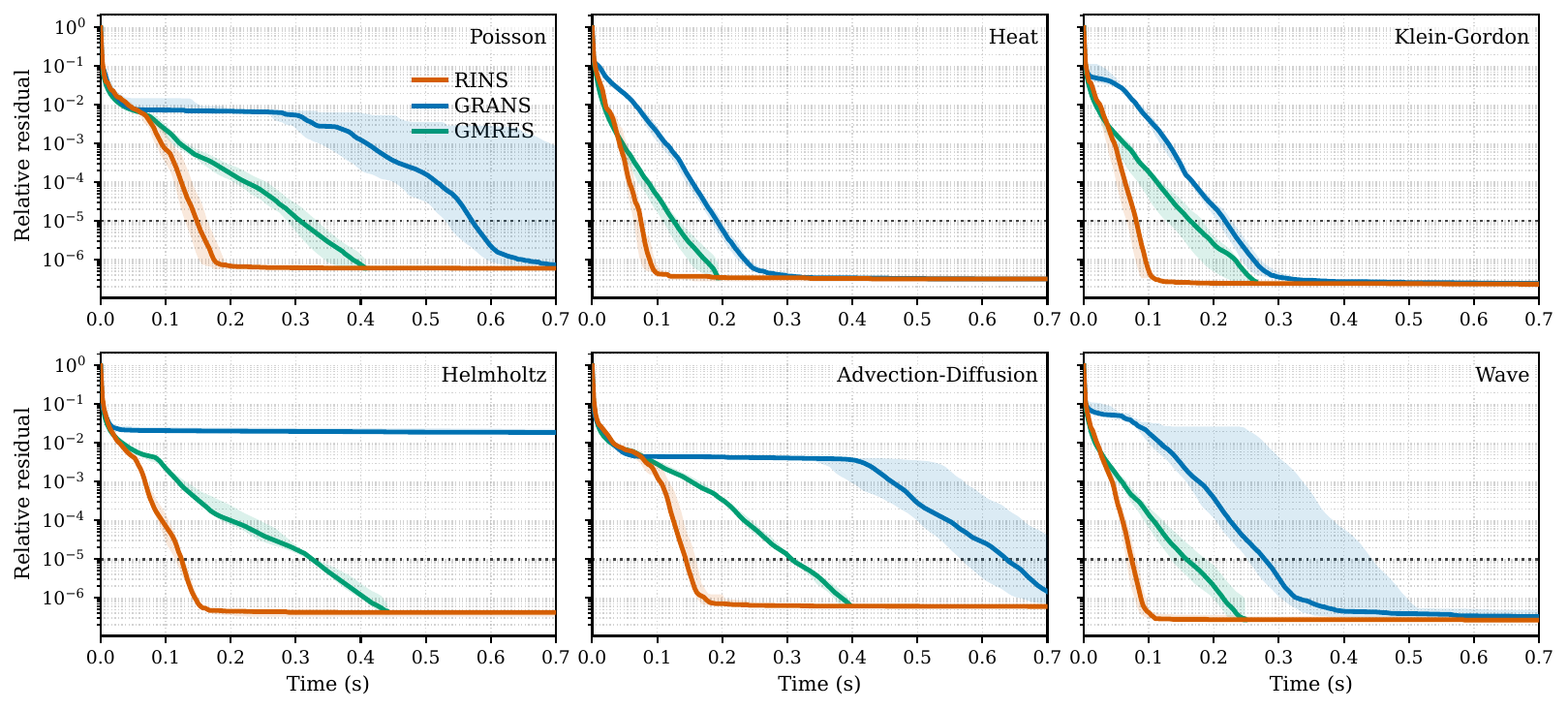}
    \caption{One-basis limit test at \(n\approx1000\). Solid curves show median relative residual versus cumulative solve time, shaded bands show interquartile ranges.}
    \label{fig:basis1_limit}
\end{figure}

We additionally run a one-basis limit test to separate the quality of each generated direction from the advantage of producing many directions per neural iteration. In the archived checkpoints used for Figure~\ref{fig:basis1_limit}, the RINS one-basis models are the gate-RINS variant with \texttt{num\_filters=1} (namely, $m$=1) and \texttt{cheb\_order=4} (namely, $L=4$). The experimental protocol is otherwise identical to the main six-PDE benchmark: each task uses the same 200 training samples and the same 10 held-out test samples, and both Static-RINS and GRANS are trained with the same three-stage schedule of 100 epochs in the first stage, 100 epochs in the second stage, and 150 epochs in the third stage. The learning rate, weight decay, checkpoint interval, loss settings, and inference protocol are also unchanged. The only architectural change is the per-iteration basis budget. In the main benchmark, both neural solvers generate 20 new basis directions at each neural iteration. In this limit test, Static-RINS and GRANS generate only one new basis direction per neural iteration. Consequently, the accumulated neural subspace dimension grows at the same rate as unrestarted GMRES.

Figure~\ref{fig:basis1_limit} shows the six \(n\approx1000\) PDE tasks. Curves report the median over the held-out test samples and shaded regions show the interquartile range; numeric markers indicate 50 and 100 generated directions or GMRES iterations. Across these tasks, Gate-RINS still decreases the residual faster than the graph-only GRANS baseline. More importantly, Gate-RINS often reaches the target residual before the accumulated subspace reaches 100 dimensions, indicating that the learned low-dimensional residual-image subspace already contains an effective correction for the solution. The same H-100 GPU timing protocol used in the main benchmark shows that, even under this one-basis-per-iteration restriction, Gate-RINS reaches the reported target residual with less iterations than GMRES on these tasks. By contrast, GRANS with only one generated direction per neural iteration fails to converge in these runs. This behavior supports the interpretation that the polynomial residual-image generator is not only benefiting from a larger per-iteration basis budget in the main benchmark; its individual generated directions are also substantially useful for the projected minimum-residual update.

\end{document}